\documentclass{aa}  

\usepackage{float}
\usepackage{graphicx}
\usepackage{txfonts}
\usepackage{amsmath,amsfonts,amssymb}
\usepackage{booktabs}
\usepackage{textcomp,gensymb}
\usepackage{changes}
\usepackage{academicons}
\usepackage{natbib,twoopt}
\usepackage[breaklinks=true]{hyperref} %% to avoid \citeads line fills
\usepackage{array,booktabs,ragged2e}
\usepackage[detect-all]{siunitx}
\usepackage[normalem]{ulem}
\newcolumntype{R}[1]{>{\RaggedLeft\arraybackslash}p{#1}}
\bibpunct{(}{)}{;}{a}{}{,} %% natbib format for A&A and ApJ
\makeatletter
\newcommandtwoopt{\citeads}[3][][]{\href{http://adsabs.harvard.edu/abs/#3}%
{\def\hyper@linkstart##1##2{}%
\let\hyper@linkend\@empty\citealp[#1][#2]{#3}}}
\newcommandtwoopt{\citepads}[3][][]{\href{http://adsabs.harvard.edu/abs/#3}%
{\def\hyper@linkstart##1##2{}%
\let\hyper@linkend\@empty\citep[#1][#2]{#3}}}
\newcommandtwoopt{\citetads}[3][][]{\href{http://adsabs.harvard.edu/abs/#3}%
{\def\hyper@linkstart##1##2{}%
\let\hyper@linkend\@empty\citet[#1][#2]{#3}}}
\newcommandtwoopt{\citeyearads}[3][][]%
{\href{http://adsabs.harvard.edu/abs/#3}
{\def\hyper@linkstart##1##2{}%
\let\hyper@linkend\@empty\citeyear[#1][#2]{#3}}}
\makeatother

\makeatletter
\renewcommand*\aa@pageof{, page \thepage{} of \pageref*{LastPage}}
\begin{document}

\title{Magnetograms underestimate even unipolar magnetic flux nearly everywhere on the solar disk}

   %\subtitle{I. Overviewing the $\kappa$-mechanism}

   \author{
J.~Sinjan\inst{1}\orcid{0000-0002-5387-636X}\thanks{\hbox{Corresponding author: J. Sinjan} \hbox{\email{sinjan@mps.mpg.de}}}
     \and
    S.K.~Solanki\inst{1}\orcid{0000-0002-3418-8449} \and 
    J.~Hirzberger\inst{1} \and
    T.L.~Riethm\"uller\inst{1} \and
    D.~Przybylski\inst{1}\orcid{0000-0003-1670-5913}
    }

   \institute{
         %1%
         Max-Planck-Institut f\"ur Sonnensystemforschung, Justus-von-Liebig-Weg 3,
         37077 G\"ottingen, Germany \\ \email{sinjan@mps.mpg.de}
         }

   \date{Received XXX XX, 2024; accepted, XXX XX, 2024}

% \abstract{}{}{}{}{}
% 5 {} token are mandatory
 
  \abstract
   {The amount of magnetic flux passing through
the solar surface is an important parameter determining solar activity and the heliospheric magnetic field. It is usually determined from line-of-sight magnetograms. 
   }
   {We aim to test the reliability of determining the line-of-sight magnetic field from a 3D MHD simulation of a unipolar region. In contrast to earlier similar studies, we consider the full solar disk, i.e. considering the full centre-to-limb variation, as well as regions with different averaged field strengths.}
   {We synthesised Stokes profiles from MURaM MHD simulations of unipolar regions with varying mean vertical magnetic flux densities, ranging from quiet Sun to active region plage. We did this for a comprehensive range of heliocentric angles: from $\mu=1$ to $\mu=0.15$, and for two commonly used photospheric spectral lines: \ion{Fe}{i}\;$6173.3$ and \ion{Fe}{i}\;$5250.2$\,\AA. The synthesised profiles were spatially foreshortened and binned to different spatial resolutions characteristic of space-based magnetographs currently in operation. 
   %No spectral degradation or other instrumental effects were taken into consideration. 
   The line-of-sight magnetic field was derived with a Milne-Eddington Inversion as well as with other commonly used methods.}
   {The inferred spatially averaged $\langle B_{LOS}\rangle$ is always lower than that present in the MHD simulations, with the exception of $\mu\approx 1$ and sufficiently high spatial resolution. It is also generally inconsistent with a linear dependence on $\mu$. Above $\mu=0.5$ the spatial resolution greatly impacts the retrieved line-of-sight magnetic field. For $\mu\leq0.5$ the retrieved $B_{LOS}$ is nearly independent of resolution, but is always lower than expected from the simulation. %We also find no change with spatial resolutions higher than $200$km. 
   These trends persist regardless of the mean vertical magnetic field in the MHD simulations and are independent of the $B_{LOS}$ retrieval method. For $\mu\leq0.5$, a larger $\langle B_{LOS}\rangle$ is inferred for the $5250.2$\,\AA\;spectral line than $6173.3\;$\AA\;, but the converse is true at higher $\mu$.}
   {The obtained results show that with high spatial resolution observations, e.g., as achieved with SO/PHI-HRT at close perihelion, the magnetic flux can be reliably retrieved at high $\mu$ values, whereas in lower resolution observations, as well as at lower $\mu$, a significant fraction of the magnetic flux is missed. The results found here raise some doubts of the reliability of determining the radial field by dividing the line-of-sight field by $\mu$ and are of considerable importance for deducing the total magnetic flux of the Sun. They may also contribute to the resolution of the open flux problem.}


   \maketitle
%
%________________________________________________________________

\section{Introduction}\label{sec:intro}

Obtaining the correct amount of magnetic flux passing through the solar surface is important for many purposes. In mixed polarity regions, the flux and its density, i.e. the field strength, determine the total magnetic energy and hence are a guide to the heating of the upper solar atmosphere. In unipolar regions associated with open magnetic field lines, such as coronal holes, the magnetic flux contributes to the heliospheric magnetic field. 

It is well known that in mixed polarity regions the measured magnetic flux density depends strongly on the spatial resolution \citep[e.g.][]{krivova2004effect,graham2009turbulent,chitta2017solar}. For unipolar fields, however, it has often been assumed that the correct amount of magnetic flux in unipolar regions is determined more or less independently of spatial resolution of the observations. 

At the same time, it has often been assumed that at least outside sunspots, most of the magnetic flux seen by full-disk magnetograms is vertical (i.e. radial), so that line-of-sight (LOS) magnetograms are sufficient to get the correct magnetic flux density, at least in unipolar magnetic regions. This assumption has been made in part due to the difficulty of retrieving a meaningful magnetic field vector outside of active regions. This in turn is because the signal in Stokes $Q$ and $U$ is typically much lower than in Stokes $V$ outside sunspots and for spectral lines in the visible spectral range, making LOS fields the most reliably deduced component of the magnetic field. The vertical component of the field can be computed by dividing the LOS magnetic field, $B_{LOS}$, by the cosine of the angle of incidence, $\theta$, i.e. the heliocentric angle, $\mu=cos(\theta)$ \citep[e.g.][]{murray1992,fligge,hagenaar}. In the following we will refer to this practice as the ``$\mu$-correction". 

Coronal holes are typical regions where unipolar magnetic fields are observed, with the most prominent being those located at the poles of the Sun around minima of solar activity cycles. Due to the geometry, retrieving the vector magnetic field is difficult there. \cite{tsuneta2008magnetic} observed large $>1$\;kG unipolar patches on the poles with Hinode SOT/SP that were predominantly radial and \cite{prabhu2020magnetic} also inferred large unipolar kG patches at the pole with the IMaX instrument on the SUNRISE balloon-borne observatory \citep[][]{solanki2010sunrise, barthol2011sunrise, martinez2011imaging, gandorfer2011filter, berkefeld2011wave} but without the need of a magnetic filling factor.

The polar magnetic field is important for a variety of reasons. For example, the polar magnetic field has been shown to be a key element to predicting the strength of the next solar cycle \citep[][]{schatten1978using,cameron2007solar,wang2009understanding}. Furthermore the fast solar wind originates from large coronal holes, i.e. from around the poles during solar minima \citep[][]{krieger1973coronal}. There is also a longstanding unsolved problem known as the open flux problem \citep[e.g. ][]{wang1995,linker2017open}. It consists of a $2-3$ factor mismatch between the open flux directly measured in-situ at $1$ au and that calculated from the magnetic flux at the solar surface in large coronal holes, and extrapolated to $1$ au. Current ideas to explain this mismatch include: missing coronal holes, open magnetic flux not in dark EUV or X-ray emission regions or an underestimation of the open magnetic flux in synoptic LOS magnetograms. This last idea strongly motivates this study: if the radial field deduced from the LOS magnetic field measured over the solar disk underestimates the actually present field, then this could contribute to solving the open flux problem.

Retrieving the vector magnetic field in coronal holes  is not straightforward, even more so when they are located at the limb, e.g. at the poles. To retrieve the complete magnetic vector full Stokes polarimetry is required. However the Stokes signals of the transverse component of the magnetic field are intrinsically lower than the longitudinal component as they are a second order effect for incomplete Zeeman splitting (which is typically the case outside sunspots for spectral lines in the visible spectrum), and therefore the signal to noise ratio is lower. First of all, coronal holes have relatively low spatially averaged field strengths, similar to quiet Sun regions, so that Stokes signals are typically small, those of Stokes $Q$ and $U$ particularly so. 

Although cancellation of Stokes $V$ profiles occurs to a much smaller degree, due to the dominance of one magnetic polarity \citep[e.g.][]{wiegelmann2004similarities}, near the limb foreshortening plays a major role as the projected area encompassed by one resolution element of the photosphere increases by $1/\mu$. The detected rays from these foreshortened areas travel a much longer path through the solar atmosphere, and when $\mu$ is low enough, rays from magnetically strong regions pass through nearby non-magnetic regions where significant absorption can take place \citep[][]{audic1991radiative,solanki1998reliability}, which further reduces the polarised Stokes parameters. The lower intensity levels near the limb also result in lower signal to noise ratios. Finally, the fact that the lines are formed higher near the limb, where the field strength in magnetic features is reduced (due to horizontal pressure balance) implies smaller Zeeman splitting which results in even lower Stokes $Q$ and $U$ profiles relative to Stokes $V$.

When observing off disc centre, the spectropolarimetric signature of magnetic features near the limb has not been well characterised. \cite{solanki1998reliability} synthesised rays utilising different flux tube models over a range of $\mu$ down to $\mu=0.2$ for a variety of spectral lines. They found that the amplitude of Stokes $V$ need not follow the $\mu$ linear dependence as expected by the $\mu$-correction. Near the limb, almost all diagnostics were greatly affected by the passage of rays through non-magnetic material, and by the width of the flux tubes in their models. They also suggested that the global magnetic flux may be underestimated near the limb. 

Hinode SOT/SP \citep[][]{tsuneta2008solar} uniquely combines high spatial resolution and polarimetric sensitivity. However, even this may not be sufficient to reliably determine the magnetic vector close to the limb \citep[][]{centeno2023limitations}. These authors  simulated polar observations by Hinode SOT/SP at $\mu=\cos(65\degree)=0.42$ and found strong biases in the retrieval of the inclination and azimuth. They point out that photon noise, projection effects, telescope spatial point spread function, spectral point spread function and the limitations of Milne-Eddington inversions (with a variable magnetic filling factor) all contribute towards the generation of these observed biases. They do, however, find that pixel-averaged quantities, such as $B_{LOS}$ when assuming a filling factor of unity, highly correlate with the MURaM simulations they used to create the synthetic observations. 

Furthermore \cite{plowman2020c} and more recently \cite{milic2024} found that with reduced spatial resolution, the mean LOS magnetic flux is not accurately retrieved: it is underestimated. \cite{plowman2020c} explained this through correlations between magnetic flux density and continuum brightness (assuming that magnetic features are comparatively dark) while \cite{milic2024} instead argue that this is due to non-linearity of the methods used to infer the magnetic field. The study by \cite{milic2024} was limited to disk centre views (i.e. $\mu=1$), which means that their results are not applicable to fields in the polar coronal holes (at least not as seen from Earth, or even from Solar Orbiter at its maximum heliolatitude; see \cite{muller2020solar}.

Due to these challenges in determining the correct magnetic flux in coronal holes, we aim to use radiative MHD simulations of unipolar photospheric regions for a comprehensive range of $\mu$ values to study how reliably the LOS field can be retrieved. In this way we not only test the $\mu$-correction to retrieve the radial magnetic field from unipolar regions away from disk centre, but also gain an idea of whether the dependence on spatial resolution of the magnetic flux obtained at disk centre is also valid closer to the limb, or if indeed magnetic flux may be generally underestimated in unipolar regions, such as coronal holes, including those near the solar limb. We do this for several spatial resolutions, mimicking those of SDO/HMI \citep[][]{scherrer2012helioseismic}, SO/PHI-HRT and FDT and lower resolution magnetographs, and for several levels of magnetic flux density, ranging from relatively quiet Sun typical of coronal holes to strong plage regions. 

In Sect.~\ref{sec:sim} we introduce the radiative magnetohydrodynamic (MURaM) simulations employed and in Sect.~\ref{sec:method} we present the method we implemented to synthesise the Stokes profiles from the MHD simulations at a range of $\mu$ values. The method to bin these profiles to different spatial resolutions is detailed in Sect.~\ref{sec:proc}. The method by which the line of sight magnetic field is inferred is outlined in Sect.~\ref{sec:cmilos}. In Sect.~\ref{sec:blos} we present the CLV of the inferred LOS magnetic field and discuss the limitations of our work in Sect.~\ref{sec:disc}. We outline our conclusions in Sect.~\ref{sec:conc}. 

\section{MURaM simulations}\label{sec:sim}
\begin{figure*}
  \centering
  \includegraphics[width=0.87\linewidth]{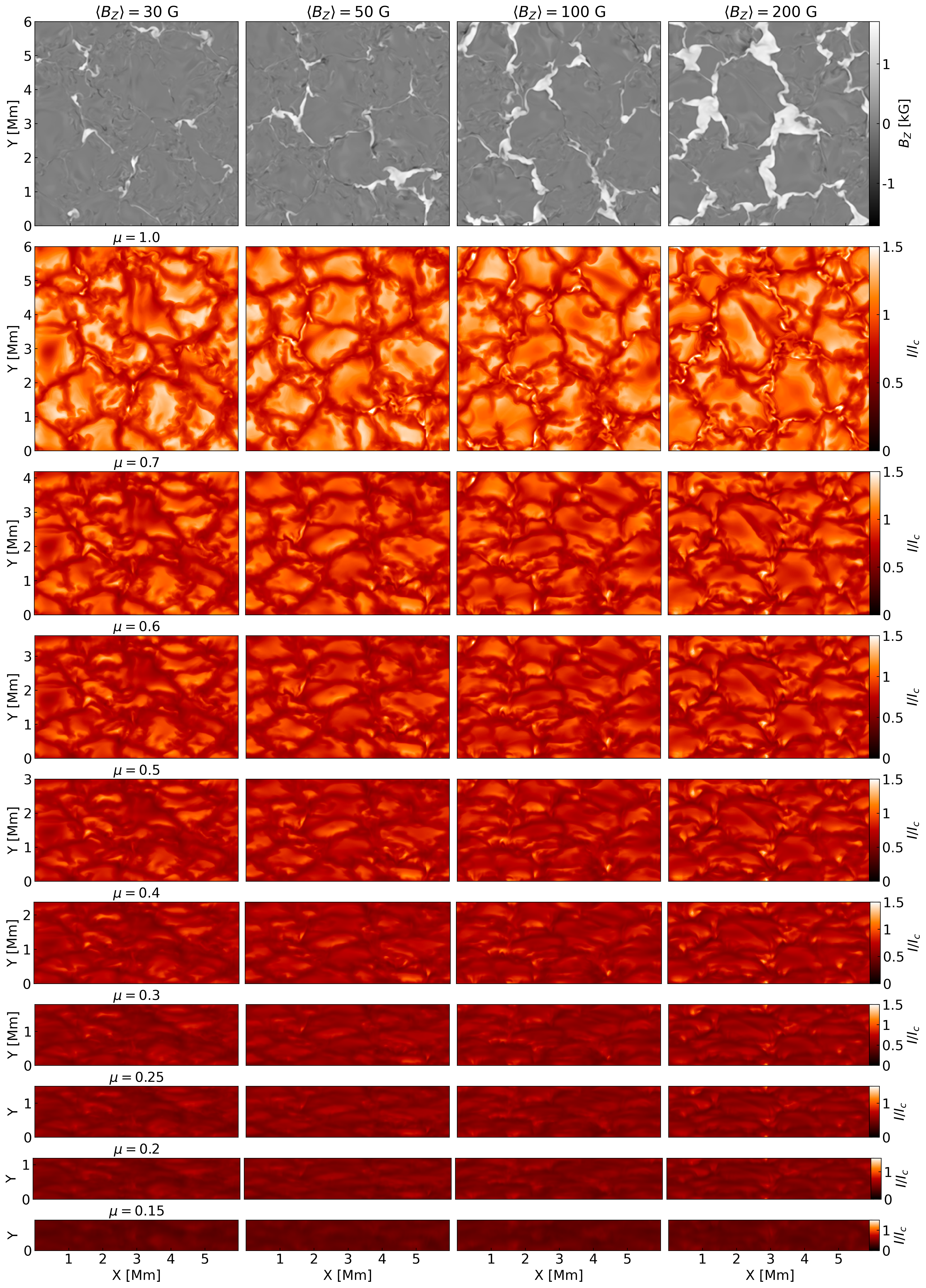}
  \caption{Top row: $B_{Z}$ in kG of the MURaM simulations at a height of $z=0.85$\;Mm above the bottom layer of the simulation box, which approximately corresponds to the visible surface layer. From left to right snapshots of simulations with different initially imposed mean vertical magnetic fields of $\langle B_{Z} \rangle=30$, $50$, $100$ and $200\;$G. Second row to bottom row: Synthetic Stokes $I/I_{c}$ maps at $d\lambda = - 0.35\;\AA$ for the \ion{Fe}{i}\; $6173.3\;\AA$ absorption line for $\mu=cos(\theta)$ decreasing from $=1.0$ to $\mu=0.15$. For $\mu<1$ the $y$ axis is foreshortened. The $I/I_{c}$ maps are saturated from $0$ to $1.5$ units of the average continuum intensity, $I_c$, at $\mu=1$. The maps at $\mu=0.9$ and $0.8$ are omitted for brevity.}\label{fig:bz_Ic}%
\end{figure*}

Three-dimensional magnetoconvection simulations of the photosphere were computed using the MURaM code \citep[][]{vogler2005simulations}. MURaM solves the MHD equations along with radiative energy transport. The radiative energy exchange rate is computed via non-grey radiative transfer under local thermodynamic equilibrium (LTE) and the equation of state takes into account partial ionisation. The simulation box has physical dimensions of $6$ x $6$ x $1.4$ Mm (x,y,z), where $z$ is the vertical axis, and has periodic horizontal boundary conditions. The box roughly covers $800$ km below the visible surface and $600$ km above. The horizontal extent of the simulation domain is sufficient in size, as at $\mu=0.15$, the extent of the foreshortened axis is $900$\;km, which equates to just over two SDO/HMI pixels, just under $9$\;SO/PHI-HRT pixels, and more than one SO/PHI-FDT pixel when SO/PHI is at perihelion. The grid size is $288 \times 288 \times 100$ cells resulting in cell sizes of $20.8$ km in the horizontal directions and $14$ km in the vertical axis. 

The simulation box has a free in- and outflow lower boundary condition and a closed top boundary, while conserving the total mass. A non-grey radiative energy transfer with four opacity bins is included. The simulation box was first initialised under hydrodynamic conditions following which a homogeneous, unipolar vertical magnetic field was introduced. The outputs of simulations with initially imposed mean vertical fields of $30, 50, 100$ and $200$\;G were included in this investigation. Analysing MHD cubes with such a range of fields allows us to determine the effect of a variety of solar environments on the determined magnetic flux density, from the weakest quiet Sun, to a strong plage region. Once the simulation had evolved to a statistically stationary state, after approximately $18$ hours of solar time, snapshots were taken every $3-7$ minutes, approximately the granule turnover time, allowing for sufficient evolution such that each snapshot is statistically independent. These simulations were produced in the initial stages of investigation for the \cite{riethmuller2014comparison} study, where they were shown to be highly consistent with observations. 

To improve the statistics, we considered $20$ snapshots each for $30$\;G and $50$\;G, $19$ for $100$\;G and $14$ for $200$\;G. Less snapshots were required for $\langle B_{Z}\rangle=200\;$G as the results did not change when more than $10$ snapshots were considered. A single snapshot for each $\langle B_{Z}\rangle$ is shown in the top row of Fig.\ref{fig:bz_Ic}. It is clear that as the mean vertical flux in the domain increases more, larger and stronger flux concentrations appear. These simulated unipolar regions represent different features on the solar surface: $\langle B_Z\rangle=30, 50$\;G represent conditions that are similar to the unipolar very quiet and the average quiet Sun, respectively (i.e. typical of coronal holes). These quiet Sun simulations well reproduce observations from the Sunrise mission \citep[][]{riethmuller2014comparison}. Meanwhile the simulations with $\langle B_Z\rangle=100$\;G represent the network field, and $\langle B_Z\rangle=200$\;G a plage region as found within active regions. Due to flux conservation and the periodic boundary conditions, the horizontally averaged (over the X-Y plane) vertical flux density, $\langle B_Z\rangle$, remains constant with geometric height in all snapshots with a numerical accuracy of $0.02$\%. This simulation setup allows us to have a known value of the flux within the simulation domain, to which we can compare the retrieved line-of-sight magnetic flux via our methods outlined in Sect.~\ref{sec:cmilos} and Sect.~\ref{ssec:blos_meths} (the actual comparisons are between spatially averaged flux density, which is equivalent). The situation does not change in principle when computing the radiation emerging at an angle form the simulation box (smaller µ), except that it is more complex because the rays often pass through multiple magnetic structures as well as the space in between \citep[see][for a description in a much more idealised version of the same geometry]{solanki1998reliability}.

\section{Spectral synthesis}\label{sec:method}

The SPINOR code \citep[][]{2000A&A...358.1109F,Frutiger_2000} was used to generate the synthetic Stokes profiles using the STOPRO routines \citep[][]{solanki_1987} for the $6173.3$\;\AA\;and $5250.2$\;\AA\;\ion{Fe}{I} absorption lines. The $6173.3$\;\AA\;line was selected as it is sampled by both the SDO/HMI and SO/PHI vector magnetographs. The $5250.2$\;\AA\ line was chosen for comparison with earlier works such as \cite{solanki1998reliability} and because both IMaX on the first two flights of Sunrise and the TuMAG instrument, a further development of the IMaX instrument, sample this line; TuMAG is scheduled to fly on the balloon-borne SUNRISE III mission in 2024 \citep[][]{herrero2022tumag}. 

For both spectral lines, the profiles were synthesised over a wavelength range of $\pm350$\;m\AA\, from the reference line core wavelength $\lambda_0$, which corresponds to a cut-off velocity of approximately $\pm17$\;km/s for $6173.3\;$\AA\;and $\pm20$\;km/s for $5250.2\;$\AA. This range was chosen to exclude contributions from nearby spectral lines, such as $6172.7$\;\AA. For the $5250.2\;$\AA\;case, neighbouring lines such as the strong Fe I  $5250.6\;$\AA\;line, were not synthesised as this would impact the retrieval of $B_{LOS}$ via standard methods such as a Milne-Eddington inversion. The spectral sampling of the synthesis for both lines is $14$\;m\AA. 

For synthesis of Stokes profiles off disc centre, the grid is adapted by the SPINOR code such that the slanted LOS (line-of-sight) becomes the new `vertical'. This is achieved by shifting the horizontal layers by $n\times D_{z}\tan(\theta)$, where $n$ is the layer index, $D_{z} = 14$\;km is the vertical grid resolution and $\theta$ is the angle of incidence i.e. the heliocentric angle. However this shift is rarely an integer value of pixels, resulting in residuals of sub-pixel shifts. To compensate, a 2D linear interpolation is performed horizontally. Furthermore, the vertical grid spacing is increased to reflect the longer path travelled through each cell: $\Delta z = D_{z}/cos(\theta)$. Next an atmosphere is built using $5000$\;\AA\ as the default reference continuum wavelength to create the optical depth scale. Each column, is converted into vertical units of $\log(\tau_{5000})$, where $\tau_{5000}$ is the optical depth at $5000$\;\AA\ continuum. When $\mu=cos(\theta)$ is very low, the increased vertical grid spacing, now in $\log(\tau_{5000})$, becomes very large, such that the vertical change in temperature can be quite large. To mitigate this, the $\log(\tau_{5000})$ grid is also interpolated in the vertical direction.

Additionally, the horizontally averaged solar surface, i.e. the layer where $\langle\tau_{5000}\rangle = 1$, is computed before the synthesis. This surface is set as the reference height layer, i.e. $n = 0$, such that it is not shifted horizontally. All other layers are shifted with respect to this reference layer. This layer is chosen to be fiducial because the continuum is formed near this layer. This choice thus mitigates unwanted artefacts from the interpolation and horizontal shifting near the height of formation of the spectral line. Stokes profiles are synthesised from $\mu=\numrange{1.0}{0.15}$ with a step size of $0.1$ in $\mu$. Below $\mu=0.3$, the step size is reduced to $0.05$. 

To improve the statistics, the profiles were synthesised from two opposite viewing directions for the full range of $\mu<1$ values. Because the simulation snapshots are not symmetric, in general the results differ somewhat for the two viewing angles. 

We now consider some of the limitations of our analysis. Firstly, the synthesis of the spectral lines did not consider any non-LTE effects, which have been shown to be significant for the Fe~{\sc i} $6173.3\;$\AA\;line \citep[][]{smitha2023non}. It is unclear, however, how large this neglection has on the results presented here. \cite{smitha2020influence} studied the effect of neglecting NLTE effects when inverting lines computed in NLTE. They found average errors in the field strength of around $10$\,G, corresponding to an average relative error of about $5\%$\ at the node with the largest signal (their Fig.~9). Although their result applies to the Fe~{\sc i}~$6301.5$ and $6302.5$~\AA\ line pair, which react somewhat differently to NLTE effects, it still suggests that uncertainty due to neglecting NLTE is considerably smaller than the effects we find here.  

Secondly, at extreme low $\mu$, the curvature of the Sun should also be considered. Here we estimate if neglecting curvature could significantly affect our results. For a ray travelling from the bottom to the top of the solar atmosphere, the change in intensity $I$ can be described by: 
\begin{equation}
    \frac{dI}{ds} = -\kappa_{\nu} + \epsilon_{\nu},
\end{equation}
where $ds$ is directed along the ray and $\kappa_{\nu}$ and $\epsilon_{\nu}$ are the absorption and emission coefficients at a wavelength $\nu$ \citep[][]{rutten}.  When assuming a plane parallel atmosphere, as we do here, we can transform $ds$ to a function of the vertical distance $z$ and $\mu$ the cosine of the angle of the ray between the vertical axis: $dz = \mu ds$:
\begin{equation}
    \mu\frac{dI}{dz} = -\kappa_{\nu} + \epsilon_{\nu}.
\end{equation}
To estimate the error by neglecting the curvature of the Sun we can transform the above into spherical coordinates, with radius $r$ and $\mu$:
\begin{equation}
    \mu\frac{\partial I}{\partial r} + \frac{1-\mu^2}{r}\frac{\partial I}{\partial \mu} = -\kappa_{\nu} + \epsilon_{\nu}.
\end{equation}
This extra term, $\frac{1-\mu^2}{r}\frac{\partial I}{\partial \mu}$, is the error we introduce by ignoring the Sun's curvature. When comparing the order of magnitudes of the first two terms we find an upper estimate for this term:
\begin{equation}
    \frac{1-\mu^2}{R_\odot} \frac{Z}{\mu},
\end{equation}
where $r$ has become $R_\odot=696\;$Mm, the solar radius, and $Z=1.4$\;Mm is the length scale of $\partial r$ of the considered simulation domain. This is an upper estimate as the key length scale for $\partial r$ is the vertical height over which the majority of the spectral line is formed, which is hundreds of km for photospheric lines rather than the entire vertical extent that we have considered here. For $\mu=0.15$, this fraction is $1.3\%$, and therefore we can safely neglect the Sun's curvature. Only below $\mu=0.05$ does this second term become significant (>$5\%$).

Maps of Stokes $I/I_{c}$, where $I_{c}$ is the average Stokes $I$ in the continuum at $\mu=1$, are shown in the native MURaM resolution from the second row to the bottom row in Fig.\ref{fig:bz_Ic} for $\mu=\numrange{1.0}{0.15}$ for different $\langle B_{Z} \rangle$. The viewing angle is such that $y=0$ is nearer to disc centre, while the top of each image is located nearer to the limb. We define this viewing direction as `positive', while viewing from the opposite side is `negative'. The CLV of Stokes $I$ can be seen, with the map becoming darker as $\mu$ approaches the limb. The profiles are spatially foreshortened to represent the projection effect.

\section{Stokes processing}\label{sec:proc}

\begin{figure*}
  \centering
  \includegraphics[width=\linewidth]{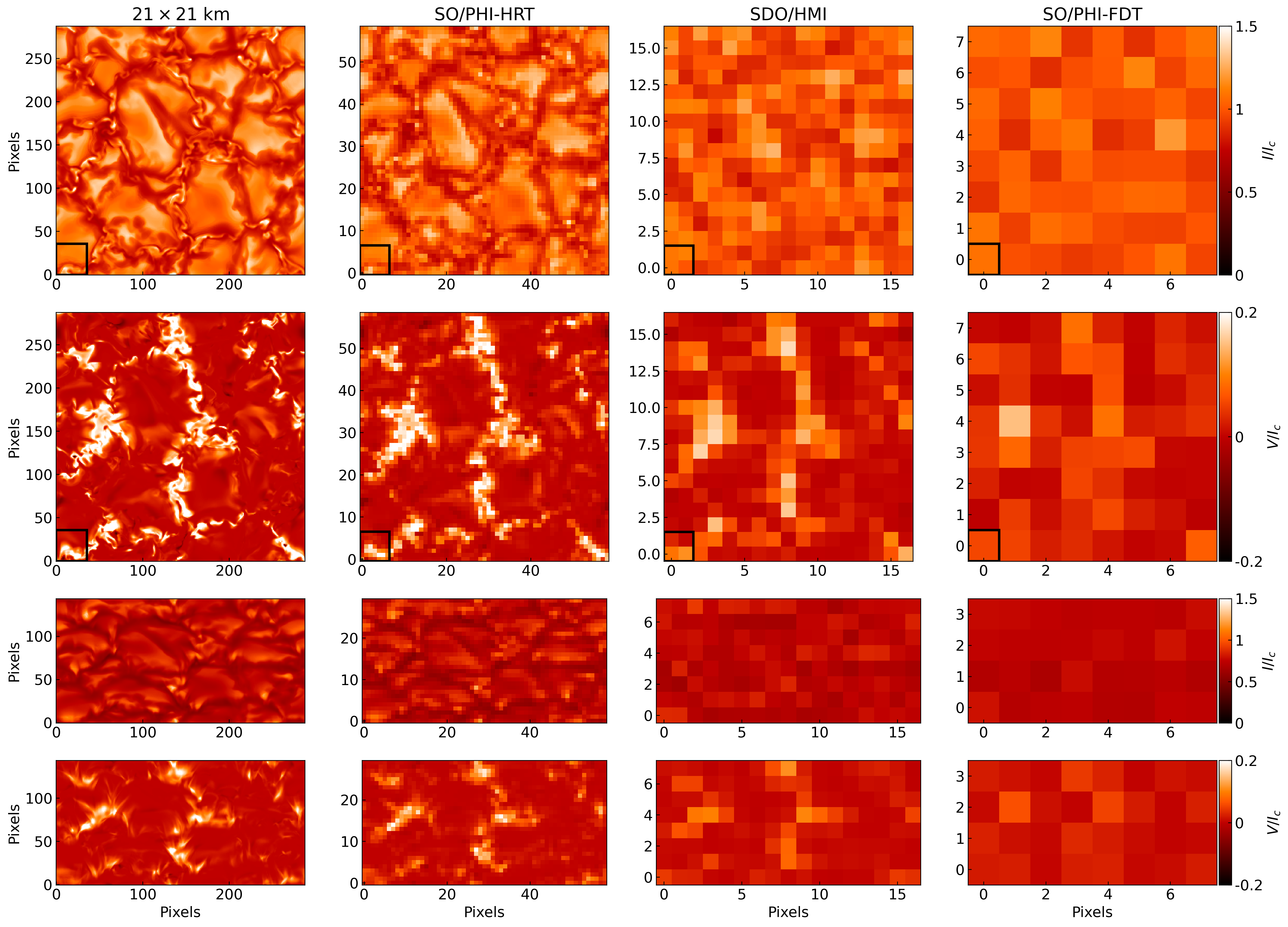}
  \caption{Maps of synthesised Stokes $I/I_c$ and $V/I_c$ profiles for the $\lambda_0=6173.3$\;\AA\;absorption line for one snapshot at $\langle B_{Z}\rangle=200\;$G at different spatial resolutions. Left to right columns: Original MURaM resolution, resolution of SO/PHI-HRT at perihelion, SDO/HMI resolution, and resolution of SO/PHI-FDT  at perihelion. Top two rows: maps at $\mu=1$. Bottom two rows: maps at $\mu=0.5$ in the `positive viewing' direction with spatial foreshortening applied. Top and third row: Stokes $I/I_c$ at $d\lambda=-0.35$\;\AA. Second and fourth row: Stokes $V/I_c$ at $d\lambda=-0.07$\;\AA. The black squares outline the region that is considered in Fig.~\ref{fig:avg_profs}.} 
  \label{fig:iv_maps}%
  \vspace{1em}
  \includegraphics[width=\linewidth]{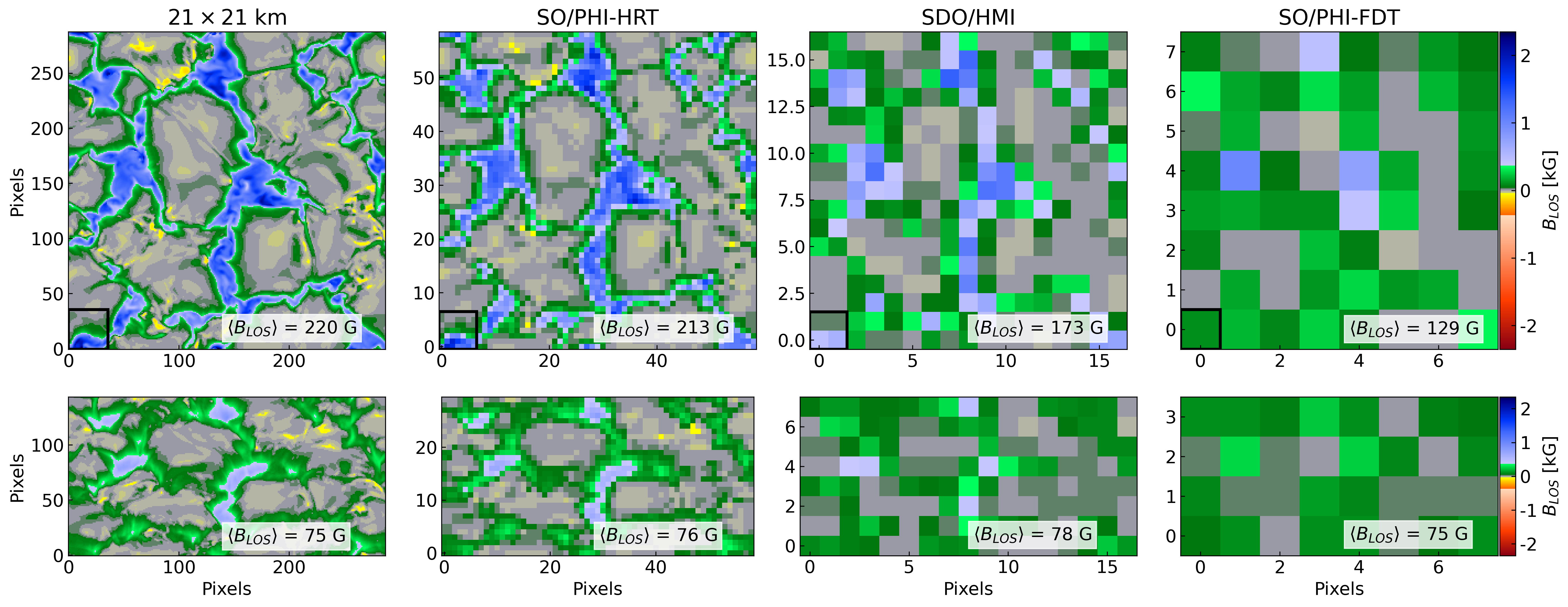}
  \caption{The $B_{LOS}$, derived via MILOS inversions, for the Stokes profiles shown in Fig.~\ref{fig:iv_maps}. Top row: $\mu=1$, bottom row: $\mu=0.5$ in the `positive viewing' direction with spatial foreshortening applied. The $B_{LOS}$ maps are shown at the original MURaM resolution, SO/PHI-HRT resolution at perihelion, SDO/HMI resolution and SO/PHI-FDT resolution, also at perihelion. The spatially averaged value of $B_{LOS}$ is inscribed at the lower-right in each panel. The black squares outline the region considered in Fig.~\ref{fig:avg_profs}.} \label{fig:blos_maps}%
\end{figure*}

The Stokes profiles, regardless of $\mu$, are synthesised on the original $288\times288$ grid of the MURaM simulations. As stated, we wish to investigate the CLV for different spatial resolutions: pixel sizes of the entire domain ($6\;$Mm), SDO/HMI ($362.5$\;km), SO/PHI-FDT ($761.5$\;km), SO/PHI-HRT ($101.5$\;km), and the original MURaM resolution ($20.8\;$km). For the two SO/PHI telescopes, these correspond to the spatial resolution achieved when Solar Orbiter is at closest perihelion: $0.28\;$au. The Hinode SOT/SP pixel size is $116\;$km, very close to that of SO/PHI-HRT at perihelion \citep[][]{tsuneta2008solar}, so that the results for SO/PHI-HRT should also be approximately valid for Hinode SOT/SP, although the spectral lines used are different. 

For the $\mu=1$ case, with no spatial foreshortening, the maps are composed of the following number of (binned) pixels (rounded to the nearest pixel): the entire domain (one pixel), SO/PHI-FDT ($8\times8$ pixels), SDO/HMI ($17\times17$ pixels), SO/PHI-HRT ($59\times59$ pixels), and the original MURaM ($288 \times288$ pixels). The binning is performed via a local mean down-sampling. 

Note that for simplicity we do not consider the spatial PSFs of the various instruments when binning to different pixel sizes. Nor do we take into account that the magnetographs sample the observed spectral line at only a very limited number of wavelength points, or the spectral PSF (i.e. filter profile).

For $\mu<1$, the pixel count, for a desired resolution, along the foreshortened $y$-axis, is calculated by dividing the foreshortened length, $6\times\mu\;\text{Mm}$, by the instrument pixel resolution. The $x$-axis pixel count remains constant independent of $\mu$. For either axis, if the desired pixel count is not an integral factor of the original $288$, the Stokes grid is first interpolated linearly in the spatial dimensions\footnote[1]{\url{https://scikit-image.org/docs/stable/api/skimage.transform.html\#skimage.transform.resize}} to a pixel count, nearest to the original $288$, that is a multiple of the desired instrument pixel count, after which the local mean down-sampling is applied. In the case of one pixel over the entire domain, the Stokes maps are spatially averaged over its entirety, regardless of the $\mu$ value. The negative viewing angles are treated as independent measurements from the positive angle, because a snapshot can present a rather different picture when observed from the two directions. 

An example of these processed Stokes profiles is displayed in Fig.~\ref{fig:iv_maps}. Here we show maps of $I/I_c$ and $V/I_c$ in the $6173.3\;$\AA\;spectral line for four different resolutions at $\mu=1$. Also displayed are the maps for $\mu=0.5$ with foreshortening applied. The $I/I_c$ maps are shown at $d\lambda=-0.35$\;\AA, i.e. basically in the continuum, while the $V/I_c$ maps are at $d\lambda=-0.07$\;\AA, i.e. in the flank of the line. It is clear that as the resolution decreases beyond that of SO/PHI-HRT the fine structure is lost. The black rectangular outline denotes the physical extent of one SO/PHI-FDT pixel at the origin at $\mu=1$. The profiles from this region are investigated in Sect.~\ref{sec:blos}.

\section{Inference of the line-of-sight magnetic field}\label{sec:cmilos}

We inferred the LOS magnetic field with an inversion code: MILOS\footnote[2]{https://gitlab.com/SOPHI1/milos} \citep[][]{suarez2007usefulness,milos}. This inversion technique solves the radiative transfer equation by assuming a Milne Eddington atmosphere. This atmosphere assumes that the physical parameters ($\mathbf{B}, v_{LOS}, \eta_{0}, \Delta\lambda_{D}$: i.e., magnetic field vector, LOS velocity, ratio of absorption coefficient at line core and continuum, and Doppler width of the spectral line) are independent of optical depth, while the source function is linear in optical depth. While we know this not to be case in the solar atmosphere, this assumption enables an analytic solution to be found, resulting in a simple and fast method to infer the physical conditions. Furthermore Milne-Eddington inversion codes have been shown to be very reliable \citep[][]{borrero2014comparison} and therefore widely used. This code is currently employed by the SO/PHI instrument \citep[][]{solanki2020polarimetric} and SDO/HMI employs a similar Milne-Eddington inversion code: VFISV \citep[][]{borrero2011vfisv}. As input the complete Stokes vector: $I,Q,U,V$ was used with the full spectral sampling of $14\;$m\AA. The LOS magnetic field, $B_{LOS}$, was determined via $B\cos(\gamma)$, where $\gamma$ is the magnetic field inclination relative to the line-of-sight. A filling factor of unity was used, the same as in the inversion routines of SO/PHI and SDO/HMI. 

When inferring the LOS magnetic field, we have neglected all other instrumental effects, such as filter profiles, discrete wavelength sampling and photon noise. The wavelength sampling and range in particular affect the retrieval of strong split profiles. To fully understand all these effects, end-to-end simulations are required, such as a study already completed for the GONG telescopes \citep[][]{plowman2020a,plowman2020b,plowman2020c}. An end-to-end simulator for SO/PHI exists \citep[SOPHISM, see][]{rodriguez2018sophism}, but results from tests across a range of $\mu$ values has not been reported yet. 

As validation of the method and its behaviour when inverting Stokes profiles with inclined lines of sight, a test case with a 1D plane-parallel atmosphere with a homogeneous vertical field of $200$\;G was used. The CLV of the inferred $B_{LOS}$ followed the expected $200\times\mu$\;G straight line very closely: the mean (vertical) separation from the $200 \times\mu$ line is $1.0\pm0.3\;$G. For more details see Appendix~\ref{append:blos_meths}.

Maps of the retrieved $B_{LOS}$, via the MILOS inversion, from the Stokes profiles underlying the images displayed in Fig.~\ref{fig:iv_maps}, are shown in Fig.~\ref{fig:blos_maps}. As illustrated by the spatially averaged $B_{LOS}$ values overlaid on each panel, one notes that the spatial average decreases with decreasing resolution at $\mu=1$. The same, however, is not the case at $\mu=0.5$, where $\langle B_{LOS}\rangle$ is approximately independent of the pixel resolution. The analysis of this behaviour is presented in Sect.~\ref{sec:blos}.

\section{$\langle B_{LOS}\rangle$ centre-to-limb variation}\label{sec:blos}
\begin{figure*}[ht]
  \centering
  \includegraphics[width=\linewidth]{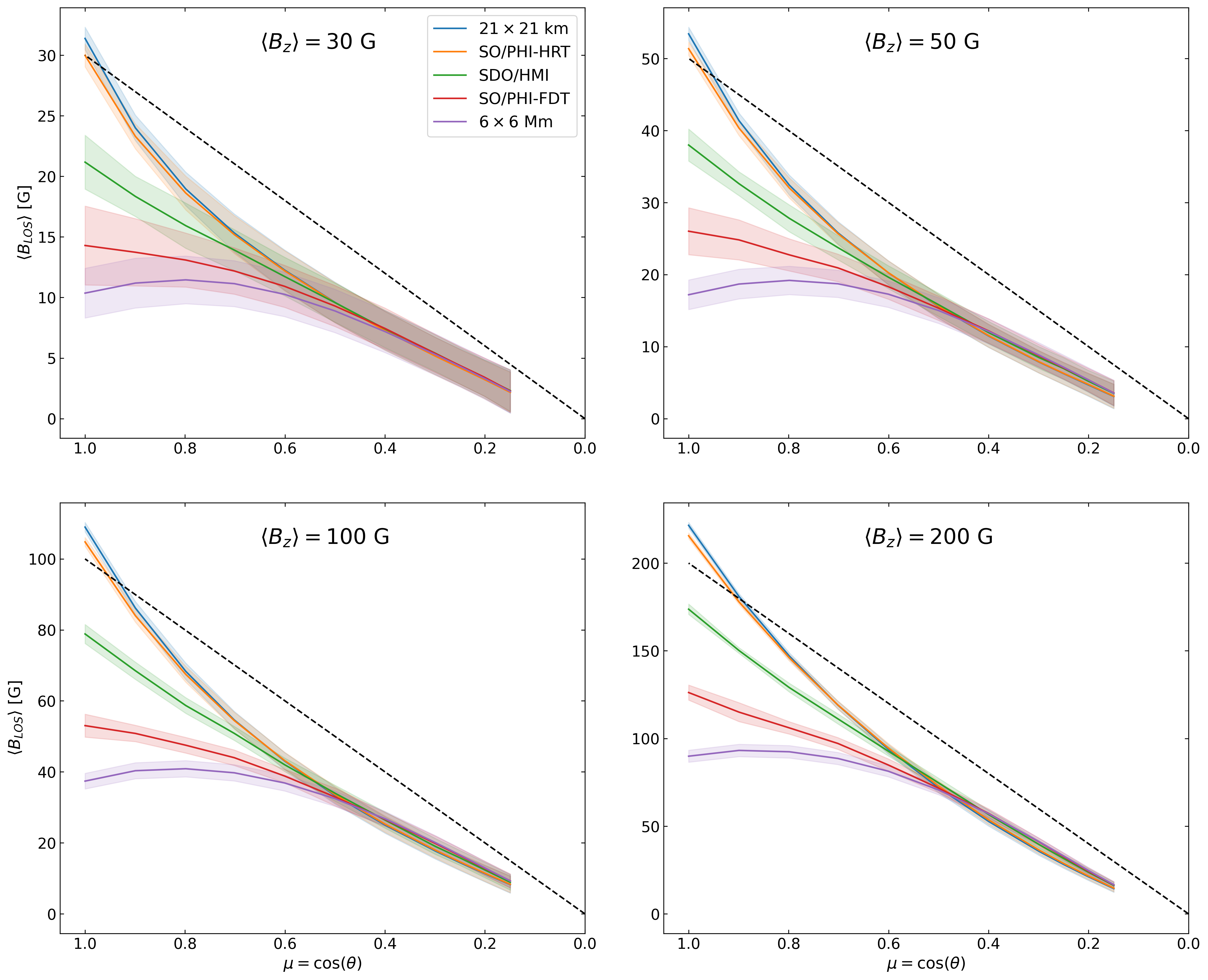}
  \caption{$\langle B_{LOS}\rangle(\mu)$ retrieved from the $\lambda_0 = 6173.3$\;\AA\;line at the five tested resolutions. From the top left panel clockwise: results for simulations with $\langle B_{Z} \rangle=30, 50, 100, 200\;$G respectively. At each $\mu$ the retrieved $\langle B_{LOS}\rangle$ quantities are averaged over all snapshots and the two employed viewing directions. The shaded regions denote the standard deviation over the range of snapshots. The legend in the upper left panel is valid for all panels. The dashed black lines indicate the expected $\mu$-dependence under the assumption that the ``ground truth'' $\langle B_Z\rangle$ is retrieved at $\mu=1$.} \label{fig:blos_clv}%
\end{figure*}
\begin{figure*}[ht!]
  \centering
  \includegraphics[width=\linewidth]{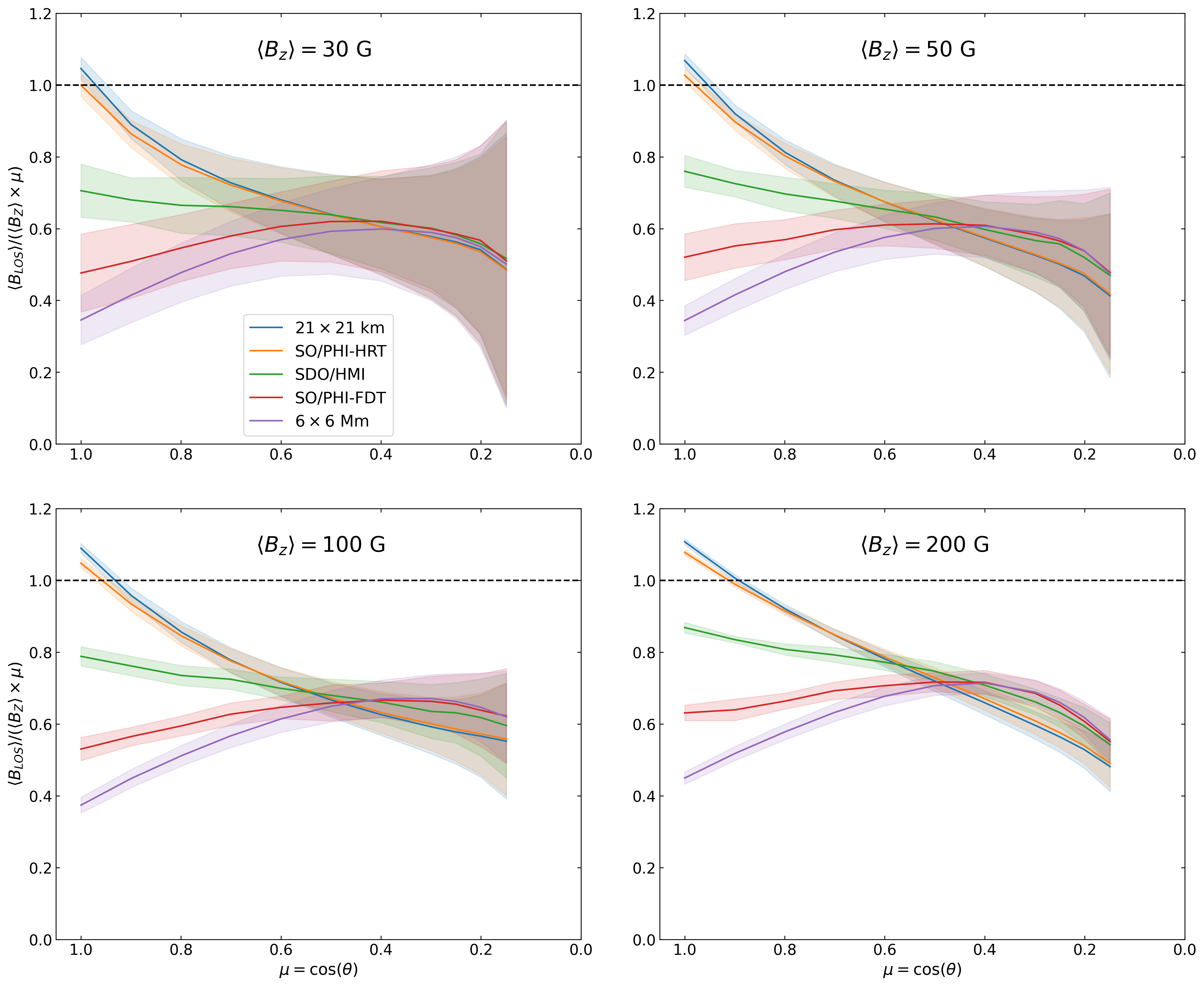}
  \caption{Same as Fig.~\ref{fig:blos_clv} but for $\langle B_{LOS}\rangle/(\langle B_{Z}\rangle\times\mu)$ instead of $\langle B_{LOS}\rangle$ plotted vs. $\mu$. The dashed black line indicates the result if the $B_{LOS}$ CLV is linear with $\mu$ and the ``ground truth'' $\langle B_{Z}\rangle$ was retrieved at $\mu=1$.} \label{fig:blos_clv_frac}%
\end{figure*}

\subsection{Fe {\sc i} $6173.3\;$\AA}\label{ssec:6173}
We determined the mean LOS magnetic field, $\langle B_{LOS}\rangle$, for the different spatial resolutions to compare how accurately the known ``ground-truth'' $\langle B_{Z}\rangle$ in the MHD simulation is retrieved. Importantly, this was done for the full range of $\mu$ values. The centre-to-limb variation of $\langle B_{LOS}\rangle$ for the $6173.3\;$\AA\;spectral line is illustrated in Fig.~\ref{fig:blos_clv}. The $4$ separate panels show that the $\langle B_{LOS}\rangle$ CLV is strikingly similar regardless of the mean vertical field strength in the MHD simulation (the main difference is that the standard deviation of points becomes smaller as the average strength of the field in a simulation box increases). The most obvious point that one draws is that the CLV, irrespective of the resolution, is almost entirely below the $\langle B_{Z}\rangle \times\mu$ curve, indicated by the black dashed lines, that is expected by the radial field assumption. This is further illustrated in Fig.~\ref{fig:blos_clv_frac}, which displays the same curves in Fig.~\ref{fig:blos_clv} but divided by the $\langle B_{Z}\rangle \times\mu$ dashed black line also shown in Fig.~\ref{fig:blos_clv}, to clearly indicate the fraction of the magnetic flux density that is underestimated.

Fig.~\ref{fig:blos_clv} and Fig.~\ref{fig:blos_clv_frac} imply that there is almost always an underestimation of flux density from unipolar regions when inferred via the $\mu$-correction, regardless of the $\mu$ value, $\langle B_{Z}\rangle$ in the MHD simulation, or spatial resolution. For $\mu>0.5$ the amount by which $\langle B_{Z}\rangle$ is underestimated increases with decreasing spatial resolution, whereas at $\mu\leq0.5$ the resolution plays a smaller role. Below $\mu=0.6$ at least $25\%$ of the flux density is `missing' when the $\mu$-correction is applied. the amount of missed flux increases to at least $40\%$ as we consider simulation snapshots with lower average flux (i.e. more representative of coronal holes). 

At or near disc centre, at SO/PHI-HRT and the native MURaM resolution, $\langle B_{LOS}\rangle$ exceeds $\langle B_Z\rangle$ by approximately $3-10\%$: at the MURaM resolution the values are: $221\pm2, 109\pm1, 53\pm1, 31\pm1\;$G for $\langle B_{Z}\rangle = 200, 100, 50, 30\,$G, respectively. This increase can be attributed to the fact that we observe the photosphere on an optical depth surface, instead of a geometric one. In regions of high magnetic flux, the plasma is evacuated such that we can `see' deeper into the Sun, into regions where the magnetic field is enhanced. \cite{schlichenmaier2023effects} presented a similar effect; \cite{milic2024} reported a $10\%$ increase for $\langle B_{Z}\rangle = 30$\;G at $\mu=1$, compared to the $\mathbf{3}\%$ increase we find, but this small difference is most likely due to the somewhat different formation heights of the $6301.5$\;\AA\;and $6302.5$\;\AA\;line pair that they investigated and $6173.2\;$\AA.

Above $\mu=0.5$, $\langle B_{LOS}\rangle$ decreases as the spatial resolution decreases, which was already visible in the $\langle B_{LOS}\rangle$ values written in Fig.~\ref{fig:blos_maps}. This appears to be caused by two effects. The first is flux cancellation: for example, in the top left quadrant of Fig.~\ref{fig:blos_maps} for $\mu=1$, there are several patches with negative $B_{LOS}$ close to positive, i.e. upward pointing, flux concentrations. When the pixel size increases, and these positive and negative flux patches are included in the same pixel, the Stokes $V$ profiles partially cancel, such that an overall lower flux density is inferred. In order to minimise this well-known and expected behaviour, we have started the simulations with purely unipolar field initial conditions. Therefore we assume that this is a minor effect compared to the other factor: as the pixel sizes increase, the strongest field regions (magnetic elements or flux tubes) are no longer adequately resolved. It is due to the non-linear behaviour of radiative transfer and the differences in properties, such as the thermal profile, between structures with and without magnetic fields, that when they are considered together in a low spatial resolution element, the inferred magnetic field does not accurately represent the true underlying physical structures.

\begin{figure}
  \includegraphics[width=0.94\linewidth]{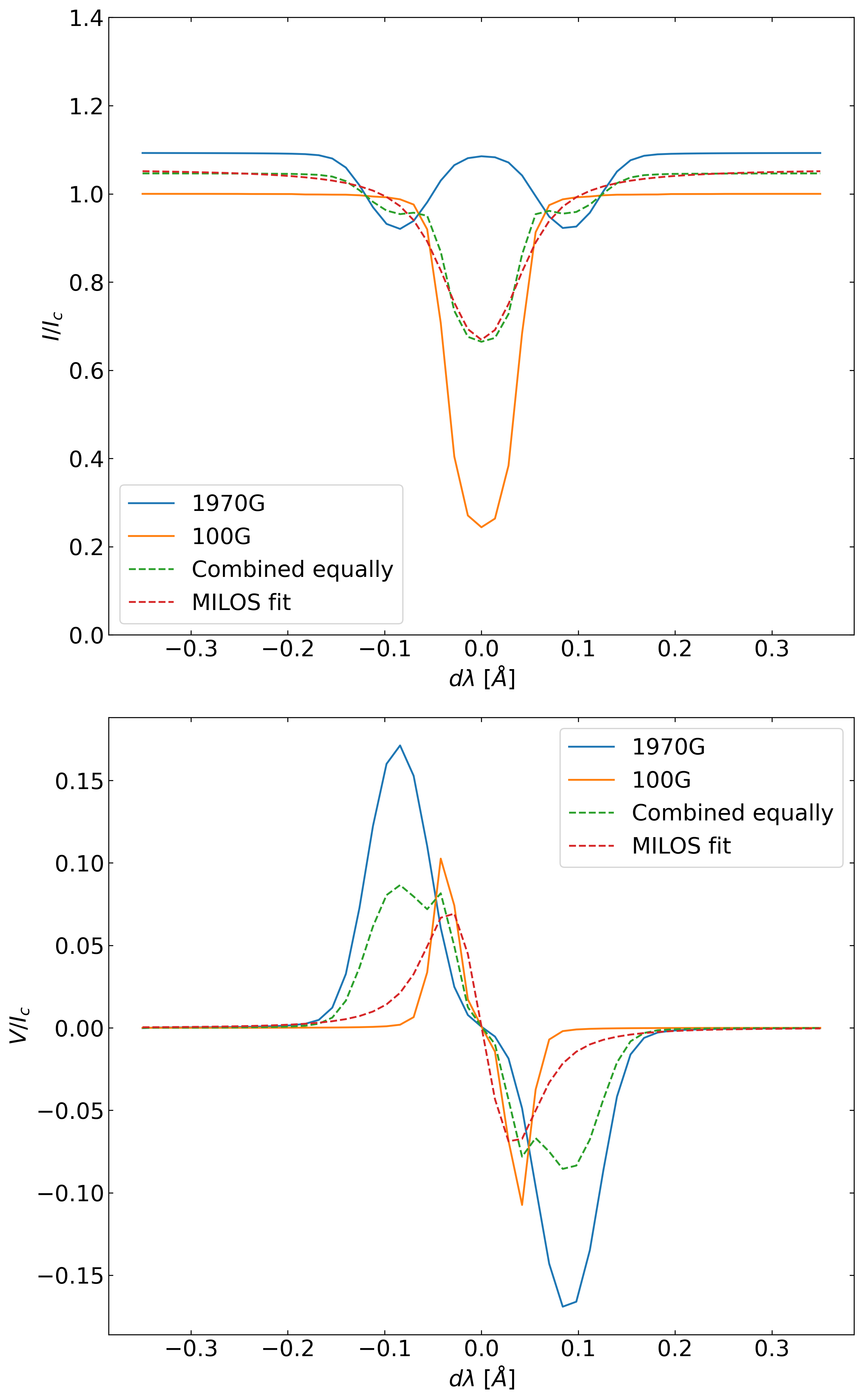}
  \caption{The combination of a weak and strong magnetic region for $\lambda_0=6173.3\;\AA$. Top panel: Stokes $I$ profiles for a $\approx2000\;$G pixel and a $100\;$G pixel. The combined profile, a simple average of the individual profiles, is depicted by the dashed green line, while the fit to the combined profile is the dashed red line. Bottom panel: same as the top but for Stokes $V$.} \label{fig:strong_weak}%
\end{figure}

An extreme example of this effect is illustrated in Fig.~\ref{fig:strong_weak}. A pixel from a snapshot with $\langle B_{Z}\rangle=200\;$G was selected with a $B_{LOS}$ of approximately $2000\;$G and low inclination ($<0.1\degree$ to the vertical) at the solar surface. The Stokes $I$ and $V$ profiles arising at $\mu=1$ from this pixel are shown in blue in the top and middle panels of Fig.~\ref{fig:strong_weak}, respectively. In orange the profile of an atmosphere with a $B_{LOS}$ of $100$\;G (and low inclination of $<0.1\degree$) at $\tau = 1$ is also depicted. Doppler shifts were removed for simplicity. When the profiles from these two atmospheres, strong and weak, are combined with equal weights (i.e. with a filling factor of $0.5$ each), the inferred $B_{LOS}$ is only $274\;$G, i.e. just under $4$ times less than the true $B_{LOS}$ in the area from which the combined profiles arise: approximately $1035\;$G. This is an extreme case, as the $1970$\;G profile is strongly split, so much so that the dips in Stokes $I$ lie almost outside the wings of the weak field Stokes $I$ profile, and hence are largely ignored by MILOS when inferring $B_{LOS}$. The fact that strong concentrations of magnetic field are typically hot leads to weaker line profiles, at least of neutral atomic lines \citep[e.g. ][]{solanki1986velocities}. This weakening also contributes to underestimating $B_{LOS}$. Another effect that helps explain the underestimate of the retrieved $B_{LOS}$ from the combined profile is the presence of Zeeman saturation for strongly split spectral lines \citep[][]{stenflo1973magnetic}. The removal of the Doppler shifts shows that this discrepancy is not a result of Doppler shifts, in agreement with the findings of \cite{milic2024}. Tests were also completed where the strong fields considered were weaker, $1000\;$G and $500\;$G. Here too the discrepancy remained although it was lessened. Using completely independent techniques, similar results for the different spatial resolutions near $\mu=1$ are found (see Sect.~\ref{ssec:blos_meths} and Appendix~\ref{append:WFACOG}\&~\ref{append:MDIWFACOG5250})

\begin{figure*}
  \centering
  \includegraphics[width=\linewidth]{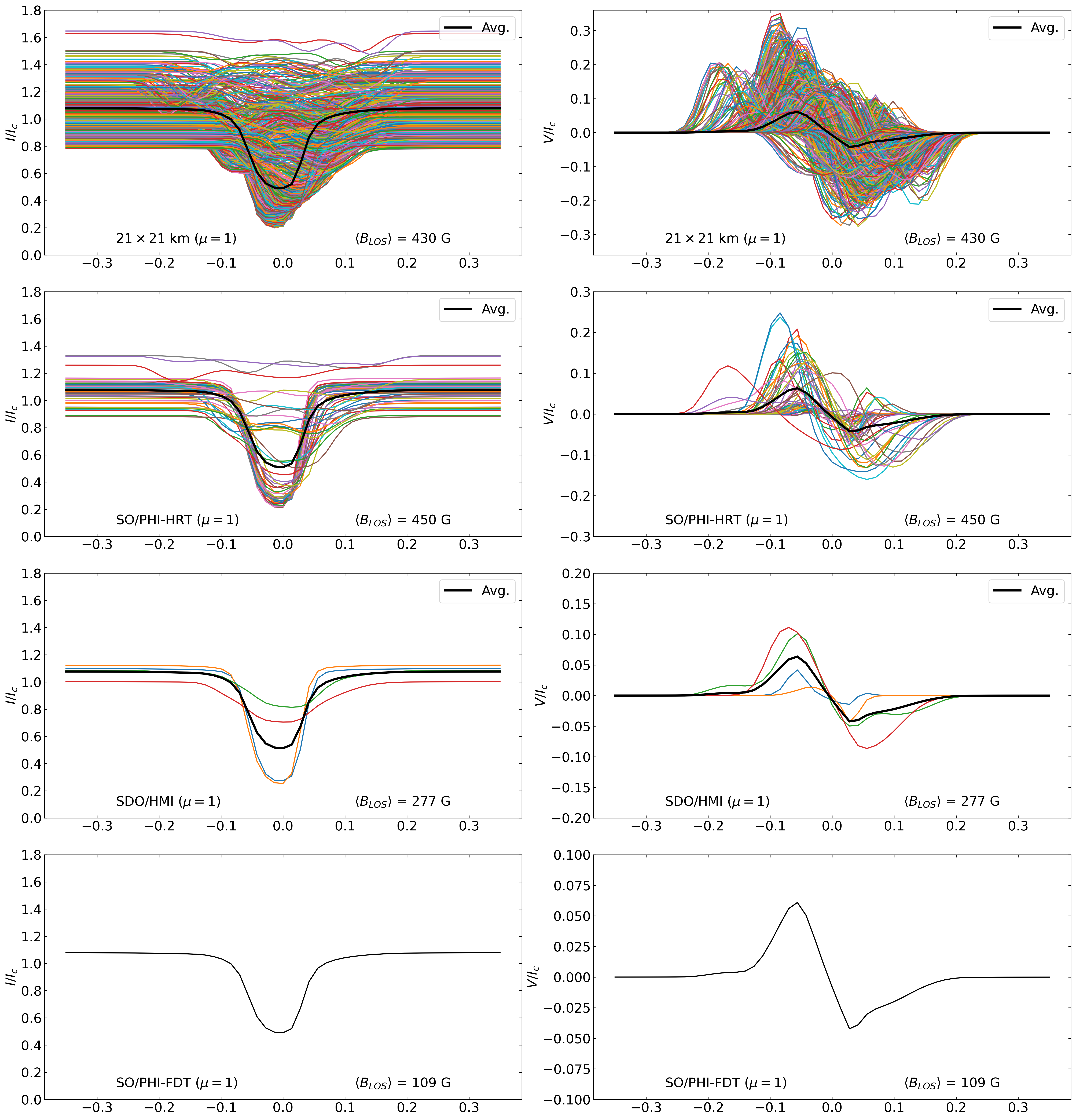}
  \caption{Stokes profiles, normalised to the spatially averaged disk centre $I_c$, from the physical region that is encompassed by the single SO/PHI-FDT pixel at $[0,0]$ at $\mu=1$, as outlined by the black square in Fig.~\ref{fig:blos_maps}. First column: Stokes $I/I_c$, second column: Stokes $V/I_c$. From top row down: the original MURaM resolution, SO/PHI-HRT resolution, SDO/HMI resolution, SO/PHI-FDT resolution. The number of considered pixel(s) in this region for the original MURaM resolution are 1296, SO/PHI-HRT: 49, SDO/HMI: 4 and SO/PHI-FDT: 1. The average line-of-sight magnetic field of the pixels in this region, $\langle B_{LOS}\rangle$, is shown for each resolution.} \label{fig:avg_profs}%
\end{figure*}

This suppression of the strong fields, and how this changes with spatial resolution is clearly observed in our results. In Fig.~\ref{fig:avg_profs} the Stokes $I$ and $V$ profiles from the region outlined in black in Fig.~\ref{fig:iv_maps} and Fig.~\ref{fig:blos_maps} is shown for $4$ different resolutions. This region is equivalent in area to one pixel at SO/PHI-FDT resolution, but $1296$ pixels at the original MURaM resolution. The profiles spatially averaged over the region are shown by solid thick black lines. If we are at original resolution, then we retrieve a $B_{LOS}$ value for the area considered that corresponds to the average of the $B_{LOS}$ values obtained from the individual pixels. This $\langle B_{LOS}\rangle$ is written in the lower right of the top panels of Fig.~\ref{fig:avg_profs}, while the averaged Stokes $I$ and $V$ profiles (in black) are much weaker and result in a significantly lower retrieved $B_{LOS}$, which is given in the lowest panels of the same figure. As indicated in Fig.~\ref{fig:avg_profs}, the spatial average of $B_{LOS}$ from the constituent pixels decreases with decreasing pixel resolution, with the $B_{LOS}$ retrieved at SO/PHI-FDT pixel resolution being a factor of $4$ lower than at the original MURaM resolution. The spatially averaged $B_Z$ in this region, close to the average height of formation, is $435\;$G\footnote[3]{The average $B_Z$ in this sub-region of the MURaM cube was found by converting geometrical height to optical depth and averaging over both the spatial ($x,y$) and 21 optical depth planes ($\log(\tau)=-2$ to $\log(\tau)=0$, with a step size of $0.1\;\log(\tau)$ and all optical depth planes weighted equally). We used response functions to the magnetic field to guide this choice. Note that this step is undertaken only for the purpose of illustration and applied only to the small sub-region of the simulation cube.} This shows that MILOS retrieves the vertical magnetic field well at high spatial resolution, while at low resolutions it does not. This difference of the average profile to that from any one pixel in the underlying area is consistent with the discussion by \cite{leka2012modeling} and a worse spatial resolution is expected to dilute the polarisation signal such that a lower magnetic field strength is inferred \cite{leka1999value,suarez2007strategy}. 

We have explained the reason why a much lower LOS flux density is inferred when the spatial resolution is low at high $\mu$ but why do the curves converge at $\mu=0.5$, and why is the LOS flux density always below the expected linear dependence, even when the spatial resolution is at the native MURaM resolution? We believe that this is a combination of mainly two factors, which have been already mentioned in Sect.~\ref{sec:intro}. Firstly, the passage of the inclined rays through both magnetic and non-magnetic regions significantly affects the inferred LOS flux density. \cite{audic1991radiative} and \cite{solanki1998reliability} described the effect on the polarised profiles by the absorption in both magnetic and non-magnetic regions, and concluded that it is highly dependent on the temperature difference between the magnetic flux tubes and the non-magnetic surroundings, and at what geometrical height the absorption takes place. Especially if the magnetic concentrations are hotter than their surroundings, which is generally the case, then the lines get more strongly absorbed in the field-free or weak-field surroundings than in the concentrations themselves. Absorption in the field-free region leads to the reduction in the strength of the polarised Stokes profiles relative to Stokes $I$. This reduces the signal of the magnetic field and mimics a weaker field, leading MILOS to underestimate $B_{LOS}$. Because this reduction is due to effects happening along each line-of-sight, it is independent of the spatial resolution of the observations. The same figure but for $\mu=0.5$ is shown in Appendix~\ref{append:prof_avg_05} and demonstrates the lack of dependence on the spatial resolution. \cite{solanki1998reliability} also found that Stokes $V$ amplitudes were more strongly reduced at $\mu<1$ for narrower flux tubes. This could explain the larger underestimation at smaller $\mu$ of regions with low $\langle B_Z\rangle$ visible in Figs.~\ref{fig:blos_clv} and \ref{fig:blos_clv_frac} as such regions tend to have narrower magnetic concentrations. Furthermore, narrow flux concentrations have a stronger tendency to be hidden behind a neighbouring granule, further reducing the polarised Stokes signals.

The second reason why we suspect the curves to converge is due to the projection effect. As $\mu$ decreases, the extent of the Sun's surface that is covered by a pixel, even at the native MURaM resolution, increases to the point that little to no strong field regions are resolved, and for those that are, the LOS component is diminished. Indeed we find that at $\mu=0.5$ at the MURaM resolution and $\langle B_Z\rangle=200\;$G, no pixels in any snapshot or viewing direction have an inferred LOS field larger than $950\;$G. 

The shaded areas in Fig.~\ref{fig:blos_clv}, colour matched with the corresponding $\langle B_{LOS}\rangle$ CLV curve, indicate one standard deviation of $\langle B_{LOS}\rangle$ from the distribution of the multiple MURaM snapshots and two viewing directions. These shaded areas reveal that the lower the $\langle B_Z\rangle$, the larger the spread in $\langle B_{LOS}\rangle$ from one snapshot to the next, and from one viewing direction to the other. This large spread in the low mean field regions could be because there are fewer and smaller flux concentrations, thus leading to more relative variation between each granule turnover time. Furthermore, as these flux concentrations are smaller the viewing geometry also has a larger impact on the inferred $\langle B_{LOS}\rangle$, as from one viewing point a flux concentration might not be visible behind a neighbouring granule, but it may well be from another. This larger variation also validates our consideration of a longer time series for the lower field strength cases.

An extreme example is found for one quiet Sun snapshot,  $\langle B_Z\rangle=30\;$G, ``observed" at $\mu=0.15$. While, as expected, a positive $\langle B_{LOS}\rangle$ was retrieved when viewing from the positive direction, a negative $\langle B_{LOS}\rangle$ is retrieved when viewing from the `negative' direction regardless of the spatial resolution, illustrating the high variability of observing the weakest quiet Sun at extreme angles. Such an apparent change in polarity may be produced by the fact that in the quiet Sun even the strong-field magnetic features are not quite vertical, so that they can be pointing towards or away from an observer observing at very small $\mu$. When there are few magnetic features in the FOV, their random inclinations need not average out.

From Fig.~\ref{fig:blos_clv_frac} we find a tendency at low $\mu$ that at high resolutions approximately $2-10\%$ lower $\langle B_{LOS}\rangle$ is inferred than at low resolutions. This is the opposite behaviour to the much more striking dependence on resolution seen at large $\mu$. The dependence on resolution increases with $\langle B_Z\rangle$ and only becomes statistically significant for $\langle B_Z\rangle=200\;$G, since at lower $\langle B_Z\rangle$ the shaded regions associated with the curves for various resolutions overlap significantly. We suspect that this difference arises as the Stokes profiles in the resolved pixels become increasingly anomalous at low $\mu$, e.g. the Stokes $V$ profiles often have more than two lobes, and do not exhibit the `normal' opposite polarity between the lobes: e.g. some profiles only have a positive signal. When MILOS then fits a profile to these anomalous profiles, $\langle B_{LOS}\rangle$ inferred from these anomalous pixels is lower than that obtained when it fits profiles corresponding to a low spatial resolution, which tend to be more normal in shape. Then the contribution of these anomalous profiles are averaged out or at least suppressed, and hence MILOS fits profiles that infer a larger $B_{LOS}$.

One more finding is the maximum of $\langle B_{LOS}\rangle$ for the lowest resolution, i.e. for the $6\times6\;$Mm case is at $\mu=0.8$ and not at $\mu=1$\ (see Fig.~\ref{fig:blos_clv}). Naively, one would expect $\langle B_{LOS}\rangle$ to decrease with $\mu$ as is the case for the other resolutions. We also find that the area and amplitude asymmetry of the Stokes $V$ profile, averaged over the entire domain, is largest at $\mu=1$, and decreases with $\mu$. It is due to this asymmetry, which is positive in both amplitude ($\approx15\%$) and area ($\approx3\%$) at $\mu=1$ (positive in the sense that the blue lobe is larger than the red lobe, see \cite{solanki1984properties} and \cite{solanki1993small} for a definition), and the inability of a Milne-Eddington inversion to fit asymmetric profiles that this maximum in $\langle B_{LOS}\rangle$ at $\mu=0.8$ exists. As the asymmetry decreases, the retrieved $B_{LOS}$ increases, but this is offset by the overall decrease in the polarisation signal as $\mu$ decreases, such that a maximum exists. This is a well known limitation of Milne-Eddington inference schemes and the resulting differences of the retrieved physical parameters with the ground truth from MHD simulations has already been presented \citep[e.g.][]{borrero2014comparison}.

We take away several key results from our work so far: the retrieved $\langle B_{LOS}\rangle$ is much lower than expected everywhere on the solar disc except for the very highest resolution observations. Close to disk centre the retrieved $\langle B_{LOS}\rangle$ depends strongly on spatial resolution, such that at low spatial resolution, the true $\langle B_Z\rangle$ is not retrieved at $\mu=1$. This result has also been reported by \citep[][]{milic2024}, although restricted strictly to $\mu=1$, who synthesised the line pair sampled by Hinode SOT/SP. This dependence on spatial resolution decreases as away from disc centre and is very weak for $\mu<0.5$. Nonetheless, also at smaller $\mu$ the  $\langle B_{LOS}\rangle$ is significantly underestimated. 

\subsection{$5250.2\;$\AA\;versus $6173.3\;$\AA\;}
\begin{figure*}
  \centering
  \includegraphics[width=0.96\linewidth]{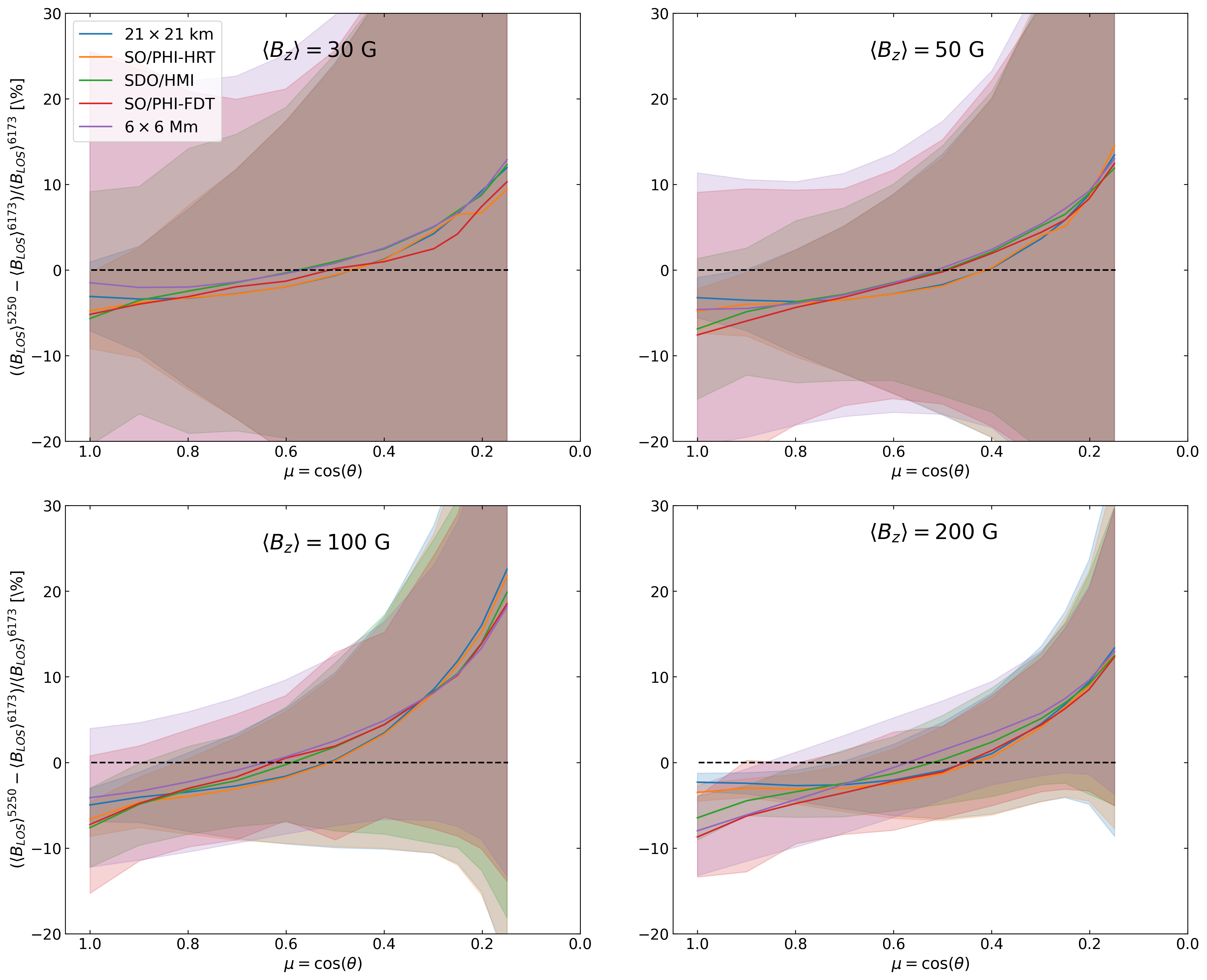}
  \caption{From the top left panel clockwise: the relative differences of $\langle B_{LOS}\rangle$, in $\%$, between the $5250.2$ and $6173.3$\;\AA\;spectral lines for $\langle B_{Z} \rangle=30,50,100,200\;$G respectively. At each $\mu$ the retrieved $\langle B_{LOS}\rangle$ quantities are averaged over all snapshots and viewing directions. The shaded regions denote the propagated standard deviation of the difference. The $y$ axis is saturated from $-20\%$ to $+30\%$. The legend in the upper left panel is valid for all panels. The dashed black line indicates the zero level.}
  \label{fig:5250_diff}
\end{figure*}

In Fig.~\ref{fig:5250_diff} the relative difference, in $\%$, between the $6173.3\;$\AA\;$\langle B_{LOS}\rangle$ CLV curves presented in Fig.~\ref{fig:blos_clv} and those for the $5250.2\;$\AA\;line are shown for the four different initial vertical magnetic fields in the MHD simulations. Across all four panels, we see a similar trend; at high $\mu$, compared to Fe~{\sc i}~$6173.3\;$\AA, up to $9\%$ lower mean LOS field is retrieved from Fe~{\sc i}~$5250.2\;$\AA. The converse is true at low $\mu$: a larger mean LOS field is inferred from Fe~{\sc i}~$5250.2\;$\AA, between $10\%$ and $23\%$ at the very lowest $\mu$ values. 

We see little dependence on the difference between the spectral lines due to the spatial resolution, only at high $\mu$ do the spatial resolutions deviate, and even in this regime the difference is only a few $\%$. It is difficult to ascertain a trend in spatial resolution in this regime, especially with the large shaded areas that indicate one standard deviation of the differences across all snapshots\footnote[4]{Using the standard error propagation formula for $y=f(a,b,...)$: $\sigma_y^{2} = (\partial y/\partial a)^{2}\sigma_a^{2} + (\partial y/\partial b)^{2}\sigma_b^{2}+....$}. Consistent with the previous figures, such as Fig.~\ref{fig:blos_clv_frac}, the statistical spread of the results across all the snapshots increases with decreasing $\langle B_Z\rangle$ in the MHD simulations.

There are various physical factors to consider here when comparing these two spectral lines, which becomes increasingly complex when inclined viewing angles are included. The Fe~{\sc i}~$5250.2\;$\AA\;spectral line has a larger absorption coefficient in the quiet Sun and it probes slightly higher in the atmosphere when viewing at disc centre. This could explain the slightly lower retrieved values at high $\mu$ \citep[][]{noda2021diagnostic}, due to the higher temperature contrast between magnetic features and the non-magnetic atmosphere at these layers. Finally, due to the much lower first excitation potential of Fe~{\sc i}~$5250.2\;$\AA, there is a large difference in temperature sensitivity of the two spectral lines. As mentioned in Sect.~\ref{ssec:6173}, when viewing at inclined angles the polarised profiles are highly affected by the temperature difference between the magnetic flux concentrations and the non-magnetic surroundings. Therefore, the difference in the temperature sensitivity of these two spectral lines may contribute to a difference in the Stokes profiles and hence the inferred LOS magnetic field. The challenge lies in disentangling all these effects to understand the behaviour we find here: to do so properly requires further study which is out of scope for this paper. Nonetheless the centre-to-limb variation of $\langle B_{LOS}\rangle$ retrieved via Fe~{\sc i}~$5250.2\;$\AA\;is qualitatively the same: $\langle B_{LOS}\rangle$ is underestimated at all $\mu$ values, at high $\mu$ it is more strongly underestimated if the spatial resolution of the synthetic observations is lower, while for $\mu <0.5$ there is little dependence on spatial resolution.

\begin{figure*}
  \centering
  \includegraphics[width=0.96\linewidth]{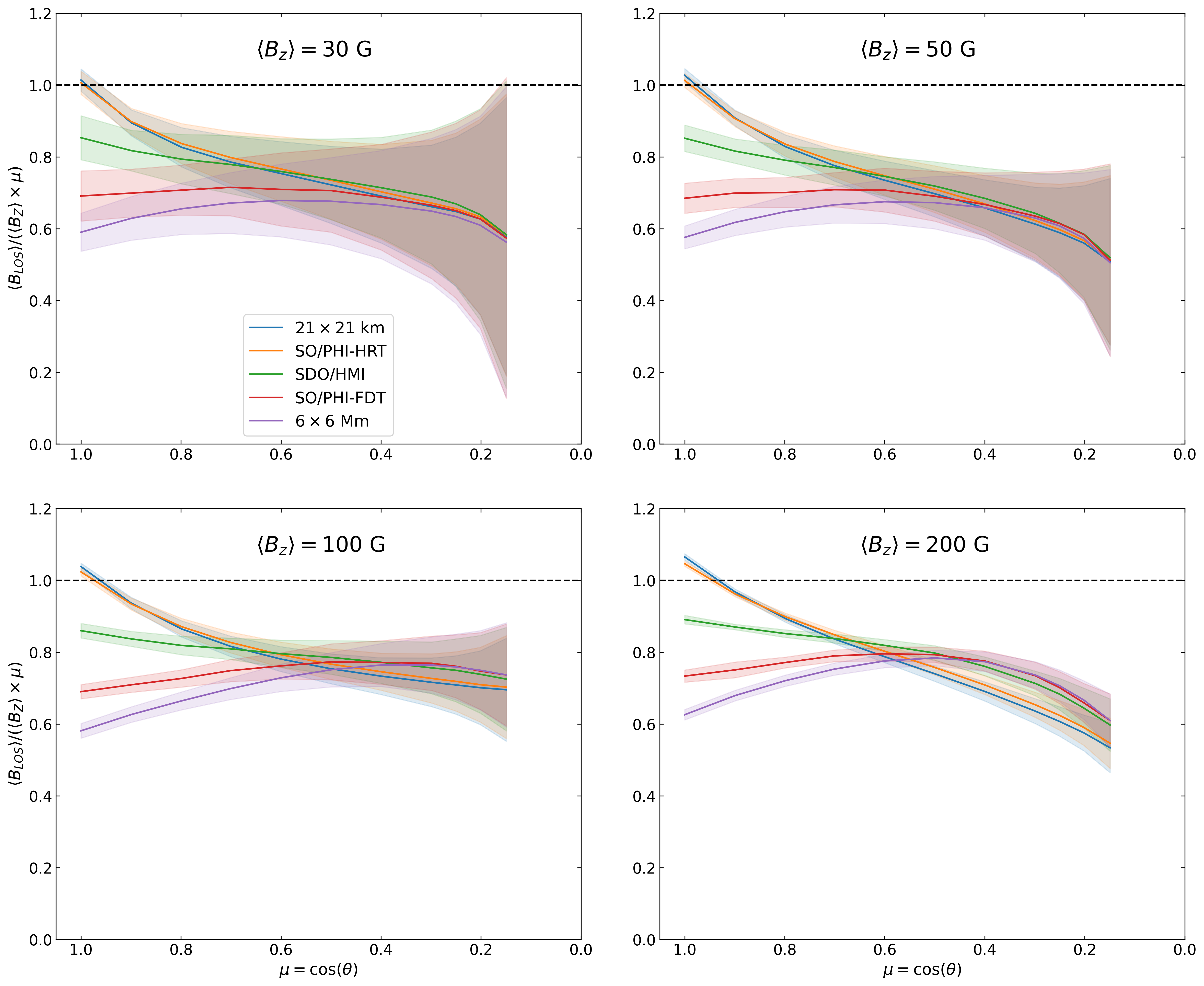}
  \caption{Same as Fig.~\ref{fig:blos_clv_frac} but derived via the MDI-like algorithm instead of with MILOS.} \label{fig:mdi_blos_clv}%%
\end{figure*}

\subsection{Other $B_{LOS}$ retrieval methods}\label{ssec:blos_meths}
To avoid that the obtained results are distorted by any bias inherent to the Milne-Eddington inversion or the MILOS code, other methods to retrieve the $B_{LOS}$ were also investigated. These included the MDI-like algorithm \citep[][]{couvidat2012line}, a Fourier Tachometer technique currently employed for the production of the LOS observables by SDO/HMI, the weak-field approximation (WFA) \citep[e.g.][]{degl2004polarization} and the centre-of-gravity (COG) method \citep[][]{rees1979line,stenflo1994}. The WFA is implemented via a linear least-squares minimisation method following \cite{gonzalez2009emergence}. These methods were validated on a one dimensional plane parallel atmosphere, as was done for MILOS, and the results of the validation are shown for $\lambda_0=6173.3\;$\AA\;in Appendix~\ref{append:blos_meths}. 

When applied to the synthetic Stokes profiles from the MURaM simulations, the WFA method retrieves results most similar to those obtained via MILOS 
inversion. However, the approximation breaks down as expected when applied to the Stokes profiles at higher spatial resolutions at high $\mu$, as the method fails to retrieve physically sensible results in pixels with the strongest fields ($B_{LOS}>1000\;$G). The COG technique and MDI-like algorithm retrieved very comparable results to each other, both of which were akin to those retrieved via MILOS: with $\langle B_{LOS}\rangle$ underestimated for all $\mu$. However the underestimation is less (i.e. the retrieved $\langle B_{LOS}\rangle$ values are larger), particularly at low resolution and high $\mu$ values. The COG technique struggled in cases where the profiles were particularly anomalous, i.e. mainly at low $\langle B_{Z}\rangle$ and low $\mu$. The results obtained by applying the WFA and COG method are given in Appendix~\ref{append:WFACOG} and ~\ref{append:MDIWFACOG5250} for the two spectral lines. 

Given this smaller underestimation of $\langle B_{LOS}\rangle$ by the COG and MDI-like technique, and that the MDI-like algorithm performed more robustly, we present the $\langle B_{LOS}\rangle$ CLV derived by the MDI-like algorithm for the $6173.3\;$\AA\;line in Fig.~\ref{fig:mdi_blos_clv}. The MDI-like algorithm infers $30-60\%$ more $\langle B_{LOS}\rangle$ compared to MILOS for the low resolution cases at high $\mu$. Nevertheless it exhibits the same trend as in Fig.~\ref{fig:blos_clv}, with a convergence of the different spatial resolution-based curves around $\mu=0.5$ and larger variance for the lower $\langle B_z\rangle$ simulations. 

As described in \cite{hoeksema2014helioseismic}, the MDI-like algorithm estimates the first Fourier coefficients, and from their phase derives the Doppler velocity (as they are proportional) for both circularly polarised components. From the difference of the Doppler velocity of these two components the line-of-sight magnetic field can be determined. This method makes several assumptions, chief among which is that the iron spectral lines are Gaussian, which we know to be only approximately correct leading to errors in the results. However, we stress that a correction for the known \ion{Fe}{I} line profile \citep[see ][]{couvidat2012line} is implemented in the SDO/HMI LOS pipeline, which we have not considered. The MDI-like method does not appear to be as sensitive to the Stokes V asymmetry as MILOS, with no maximum for the $6\times6\;$Mm pixel resolution at $\mu=0.8$. 

\section{Discussion}\label{sec:disc}

In this section we discuss implications regarding the work presented in this study. One of our main results is that, even for relatively vertical unipolar fields organised in flux tubes, we infer the wrong magnetic flux density at disc centre, with the derived flux density being clearly too low for synthetic observations having spatial resolutions worse than 200 km on the Sun. 

This result has also been reported by \cite{milic2024}, but our work is  complementary and goes beyond what those authors have presented. We can confirm this result for multiple flux densities ranging from the very quiet Sun to a reasonably strong active region plage, while \cite{milic2024} only investigated one flux density. We find very similar behaviour across all tested flux densities, but with less variance between different MHD simulation snapshots at higher flux densities. Additionally, unlike \cite{milic2024}, we did not solely rely upon the retrieval of the line-of-sight magnetic field by a Milne-Eddington inversion. We also investigated three additional widely used techniques, the weak-field approximation (linear least-squares minimisation implementation), an MDI-like algorithm and the centre-of-gravity technique. The weak-field approximation generally gave results closest to the Milne-Eddington inversion, but, unsurprisingly, broke down at locations where strong fields were spatially resolved. With the MDI-like algorithm and centre-of-gravity technique the inferred flux density was found to be somewhat closer to the ground truth, but it was still significantly underestimated at low spatial resolutions. Finally, we demonstrated that this result holds also for two widely used spectral lines Fe~{\sc i}~$6173.3\;$\AA\ (used e.g. in SDO/HMI and SO/PHI) and Fe~{\sc i}~$5250.2\;$\AA\ (used e.g. by Mt. Wilson and Kitt Peak observatories and the IMaX, TuMag instruments on the Sunrise balloon-borne observatory), which were not investigated by \cite{milic2024}. 

Another key result is that we retrieve LOS magnetic fields that are much lower than expected by the radial field assumption also off disc centre, in fact at any angle down to $\mu=0.15$. The centre-of-gravity technique struggled at low $\mu$ due to the increasingly anomalous profiles, while the other methods were more robust. These results have strong implications for total flux measurements, of both open and closed magnetic field regions, irrespective of the technique used to infer the magnetic field. The Fe~{\sc i}~$5250.2\;$\AA\;and Fe~{\sc i}~$6173.3\;$\AA\;spectral lines have both been extensively employed by past and present observatories. Hence our results support the case that a significant amount of magnetic flux is missed by Zeeman-effect based flux measurements. This includes magnetic fluxes reported by long term monitoring programs such as Mount Wilson and Kitt Peak \citep[][]{arge2002two,wallace2019estimating}, which have relatively low spatial resolution, but also more recent observations by, e.g., the widely used SDO/HMI instrument. This applies also to regions such as unipolar plage within active regions and coronal holes.

In the final paper of a series of papers that describe the end-to-end simulation of the GONG telescopes, \cite{plowman2020c} report that GONG too underestimates the magnetic flux, although they sample a different line: Ni~{\sc i}~$6768\;$\AA. As mentioned in Sect.~\ref{sec:intro}, they explain their results through convective blue-shift, in contrast with our argument at disc centre of the non-linear behaviour of the spectral lines in the combination of structures with and without magnetic fields (thermal effects, Zeeman saturation, and details of the fitting of complex line profiles), which is qualitatively in agreement with \cite{milic2024}. Near the limb, we attribute the decrease to the strong absorption in the nearly field-free gas between magnetic flux concentrations through which each ray passes, even those that pierce the magnetic concentrations, an effect proposed by \citet{audic1991radiative} and \citet{solanki1998reliability}. Since the latter effect acts along individual rays, it is expected to be almost independent of spatial resolution, which is what our computations show.

Together, our work, that of \cite{plowman2020c} and of \cite{milic2024} suggest that the underestimation of magnetic flux due to low spatial resolution near disc centre is independent of the spectral line sampled, although this needs to be confirmed in future work. Whether the difference in result between our work and that of \cite{plowman2020c} is partly due to the different spectral lines studied, needs further analysis.

We have restricted ourselves to unipolar regions for two reasons. Firstly, we expected the spatial resolution of observations to play a much smaller role for unipolar fields than for mixed polarity fields, where flux cancellation is known to play a large role in reducing the deduced magnetic flux for lower spatial resolution (at all $\mu$). Secondly, unipolar fields are associated with coronal holes and open magnetic flux. Hence our study may shed some light on whether effects such as Zeeman saturation, thermal line weakening in magnetic features, saturation effects due to radiative transfer along rays passing through both magnetic concentrations and the nearly field-free gas in between, and spatial resolution effects could help explain the mismatch between the open magnetic flux seen in coronal holes and the heliospheric magnetic flux measured in situ \citep[][]{linker2017open}. 

Since at activity minimum the coronal holes are mainly found in the polar regions, and hence are located close to the solar limb as seen from Earth, it is important to not just restrict such a study to $\mu=1$, but to consider the full centre-to-limb-variation. Here we do not aim to provide a quantitative estimate of how much of the observed mismatch can be explained by the effects we have found. That will be the topic of a follow-up investigations. 

Combined observations by the SO/PHI instrument and SDO/HMI may provide a route to observationally test the results of the present study for $\mu<1$. This will also be the subject of a future study.

\section{Conclusions}\label{sec:conc}

We have used MHD simulations of unipolar magnetic regions representing a range of solar features from the quiet Sun to active region plage in the photosphere to test how reliably various methods retrieve the LOS magnetic field from Stokes profiles. To this end, Stokes profiles of two commonly used spectral lines, Fe~{\sc i}~$6173.3\;$\AA\;and Fe~{\sc i}~$5250.2\;$\AA, were synthesised in sets of simulation snapshots covering a range of magnetic flux densities. The synthesis was repeated for a range of viewing angles corresponding to synthetic observations ranging from solar disc centre down to $\mu=0.15$. We have binned these Stokes profiles to different spatial resolutions and inferred the line of sight magnetic field using a variety of techniques. We have shown that the mean line-of-sight magnetic field is underestimated in nearly all cases at all $\mu$. The only exception is close to disc centre, where high resolution observations that resolve the strongest field  regions, which return roughly the correct averaged magnetic flux density. However, below $\mu\approx 0.9$ the synthetic observations underestimate the magnetic flux independently of the spatial resolution. We also find that for $\mu$ less than $\approx 0.5$ the retrieved averaged flux density is nearly independent of spatial resolution of the synthetic observations and is always well below the expected value. This underestimation was consistently found regardless of the inference method used.  

Our results also suggest the inferred line-of-sight flux density does not depend linearly on $\mu$, but rather shows a more complex dependence and lies too low. Hence, when applying the $\mu$-correction (i.e. dividing $B_{LOS}$ by $\mu$), there is a significant discrepancy between the inferred radial magnetic field and the true radial field in the simulation. This discrepancy is enhanced with low resolution for $\mu>0.5$, as the mixing of the strongest flux regions with neighbouring weak flux regions results in a much lower than expected inferred flux density (due to effects such as thermal line weakening in the magnetic features and Zeeman saturation, etc.). 

At or below $\mu=0.5$ there is little to no variance in the inferred mean magnetic flux density with spatial resolution. This is due to radiative transfer effects that act on individual light rays passing through both magnetic and non-magnetic regions. The large projection effect overwhelming any possibility of resolving the strongest field features may also play a role. 

We do find non-negligible differences between the two spectral lines, Fe~{\sc i}~ $6173.3\;$\AA\;and $5250.2\;$\AA, considered in this study. At high $\mu$, an up to $9\%$ lower mean line-of-sight flux density is inferred from Fe~{\sc i}~$5250.2\;$\AA\ relative to Fe~{\sc i}~$6173.3\;$\AA. The difference is particularly striking at low spatial resolution. At low $\mu$, however, the opposite is true where instead an up to $23\%$ larger mean flux density is inferred from Fe~{\sc i}~$5250.2\;$\AA. Nonetheless, this line significantly underestimates $B_{LOS}$ everywhere as well. 

These results were consistently found for all $\mu$ values irrespective of the exact employed magnetic field inference technique and of the mean vertical flux density of the simulated region. A small dependence on the mean flux density in the MHD simulations is found, with the synthetic  measurements applied to simulations with lower magnetic flux typically falling short more severely. All this suggests that the main results are robust, and hence have significant implications for a broad range of reported solar observations.

The results presented here have a potential to contribute substantially to the resolution of the open flux problem \citep[][]{linker2017open}, as the magnetic flux is underestimated everywhere on the solar disk in unipolar fields such as those underlying coronal holes, even at the spatial resolution of SDO/HMI. Turning this into estimates of the amount of "missing" magnetic flux will be the subject of a follow-up paper.

\begin{acknowledgements}
       J.S. is funded by the International Max Planck Research School (IMPRS) for Solar System Science at the University of G\"ottingen. In this work we used sunpy v4.1.7 and astropy v5.2.2 python packages. This project has received funding from the European Research Council (ERC) under the European Union's Horizon 2020 research and innovation programme (grant agreement No. 101097844 — project WINSUN). 
\end{acknowledgements}


\bibliographystyle{aa}
\bibliography{bibfile.bib}

\appendix
\onecolumn

\section{Validation of $B_{LOS}$ retrieval methods for a 1D atmosphere for all $\mu$}\label{append:blos_meths}
\begin{figure*}[hb]
  \centering
  \includegraphics[width=\linewidth]{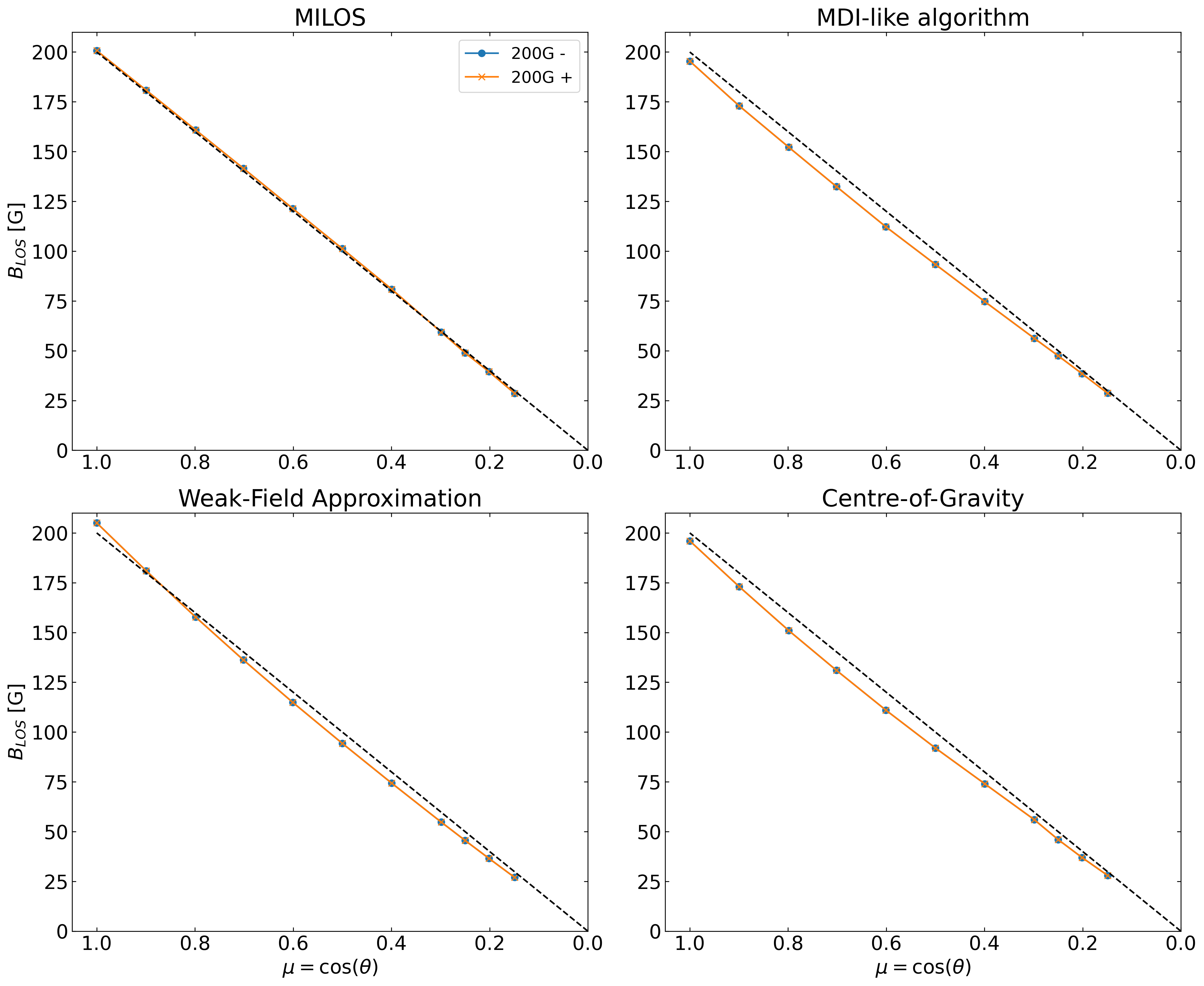} %trim={0 0 0 0},clip,
  \caption{$B_{LOS}$ CLV retrieved through $4$ different methods when applied to a 1D plane parallel atmosphere, where $B_{Z}=200\;$G everywhere, for both positive and negative viewing directions. Clockwise from top left: MILOS, MDI-like algorithm, linear least squares weak-field approximation and the centre-of-gravity method.} \label{fig:blos_meths_ppa}%
\end{figure*}

The SPINOR synthesis and $B_{LOS}$ retrieval methods were validated by first generating Stokes profiles for a 1D plane parallel atmosphere (PPA), where $B_{X},B_{Y}=0$ and $B_{Z}=200\;$G everywhere. The atmosphere was calculated by averaging a snapshot of the $B_{Z}=200\;$G MURaM simulation \citep[][]{riethmuller2014comparison}, over the horizontal ($x,y$) dimensions, and setting $v_x,v_y=0\;$km/s. Under these conditions one expects to retrieve exactly $B_{LOS}(\mu) = 200\times\mu$. As shown in Fig. \ref{fig:blos_meths_ppa} all tested methods retrieve the input field well, with MILOS doing somewhat better than the other techniques. The average (vertical) separation of the retrieved CLV from the expected $200\times\mu$ for the $4$ methods are as follows: MILOS $1.0\pm0.3\;$G, MDI-like algorithm $5\pm2\;G$, WFA $4\pm1\;$G and COG $6\pm2\;$G. All panels demonstrate that the SPINOR synthesis code generated the negative angles correctly as they lie on top of the positive viewing angle curves. This proves that any difference between negative and positive viewing direction presented in this study are purely the result of the anisotropic nature of the features in the generated photospheres. This figure also validates the correct implementation of methods and also demonstrates that any deviation from the dashed line arises from radiative transfer effects.

\newpage

\section{Stokes Profiles at $\mu=0.5$}\label{append:prof_avg_05}
At $\mu=0.5$, this region (the SO/PHI-FDT pixel at the origin) is no longer the same physical area as that considered in Fig.~\ref{fig:avg_profs}. This is due to the foreshortening, which means that at $\mu=0.5$, a larger physical region is sampled. At this $mu$ value, the region is twice as large in the $y$ dimension, equal to two SO/PHI-FDT pixels at disc centre. The spatially averaged magnetic field along the line of sight in the MURaM simulation from this larger area, between $\log(\tau)=-2$ and $0$, is $54\;$G. 
\begin{figure*}[hb]
  \centering
  \includegraphics[width=\linewidth]{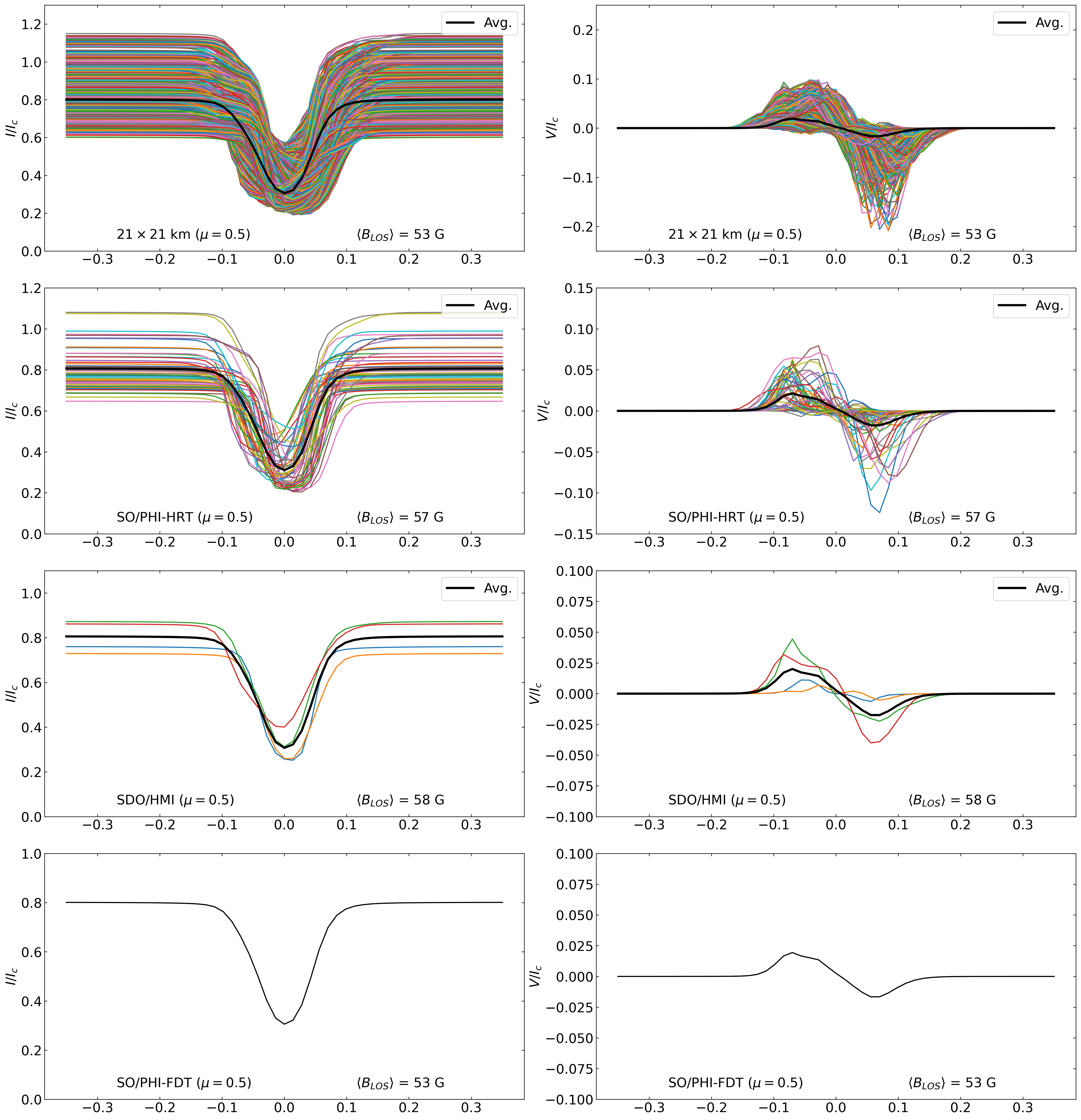}
  \caption{Same as Fig.~\ref{fig:avg_profs} but for $\mu=0.5$. The physical region encompassed by this region, due to the foreshortening, however is not the same. See text for details.} \label{fig:avg_profs_05}%
\end{figure*}

\newpage
\section{$\langle B_{LOS}\rangle$ centre-to-limb variation for Fe~{\sc i}~$6173.3\;$\AA\;derived by the weak-field approximation and centre-of-gravity technique.}\label{append:WFACOG}

\begin{figure*}[hb]
  \centering
  \includegraphics[width=\linewidth]{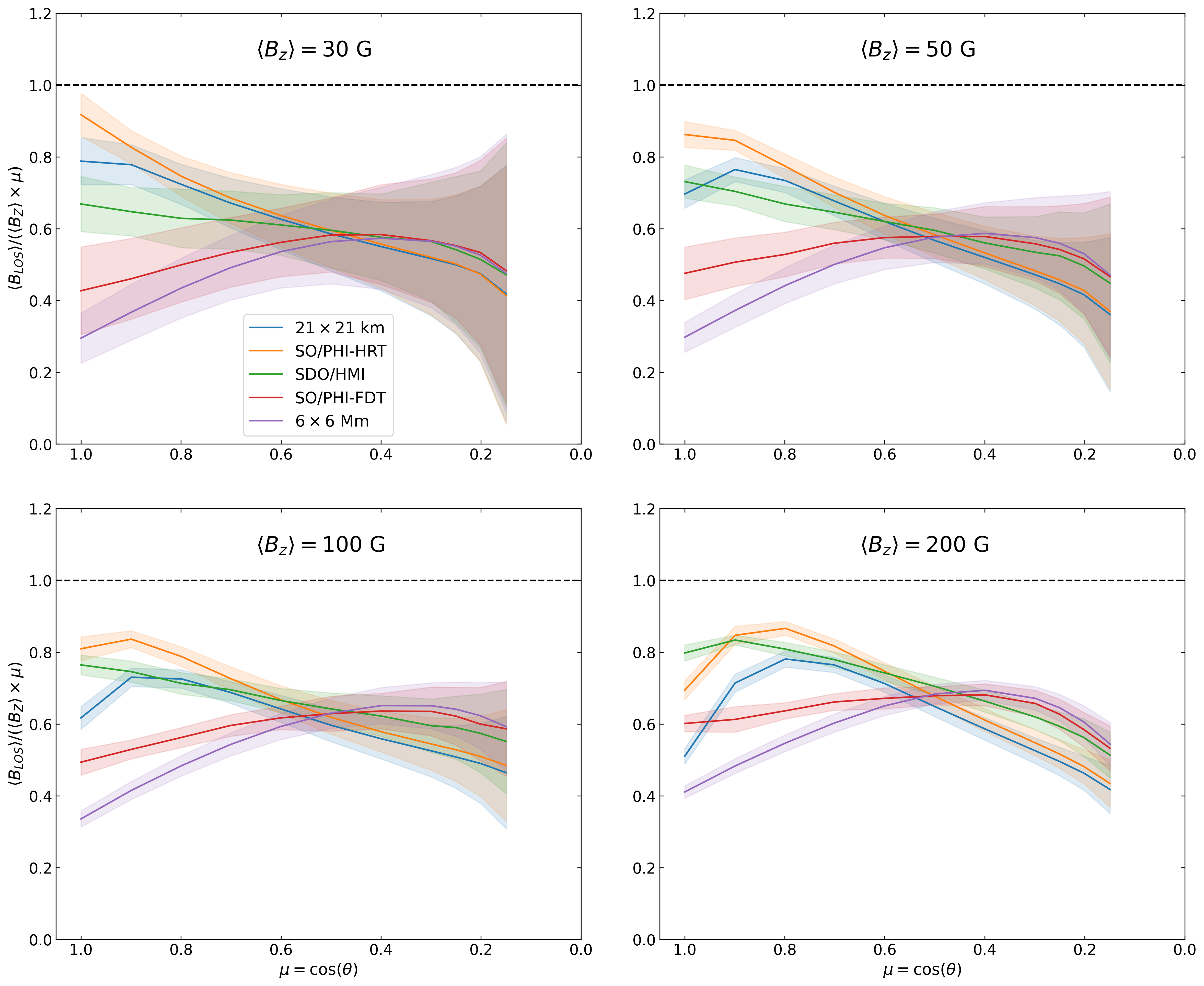}
  \caption{Same as Fig.~\ref{fig:blos_clv_frac} but $B_{LOS}$ is inferred by the weak-field approximation.} \label{fig:wfa_clv}%
\end{figure*}

\begin{figure*}
  \centering
  \includegraphics[width=\linewidth]{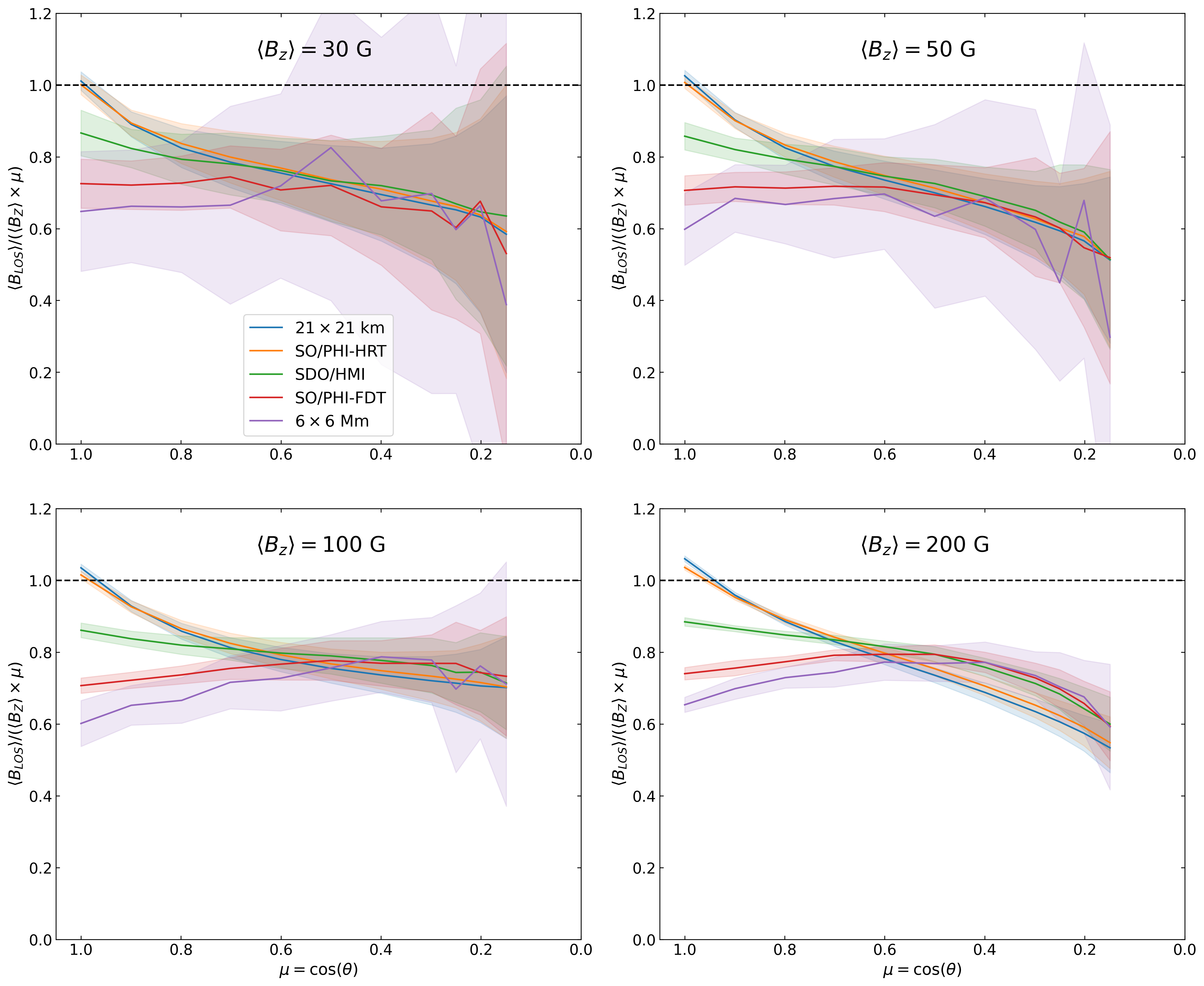}
  \caption{Same as Fig.~\ref{fig:blos_clv_frac} but $B_{LOS}$ is inferred by the centre-of-gravity technique.} \label{fig:cog_clv}%
\end{figure*}

\newpage
\newpage

\section{$\langle B_{LOS}\rangle$ centre-to-limb variation for Fe~{\sc i}~$5250.2\;$\AA\;derived by the MDI-like algorithm, weak-field approximation and centre-of-gravity technique.}\label{append:MDIWFACOG5250}

\begin{figure*}[h]
  \centering
  \includegraphics[width=\linewidth]{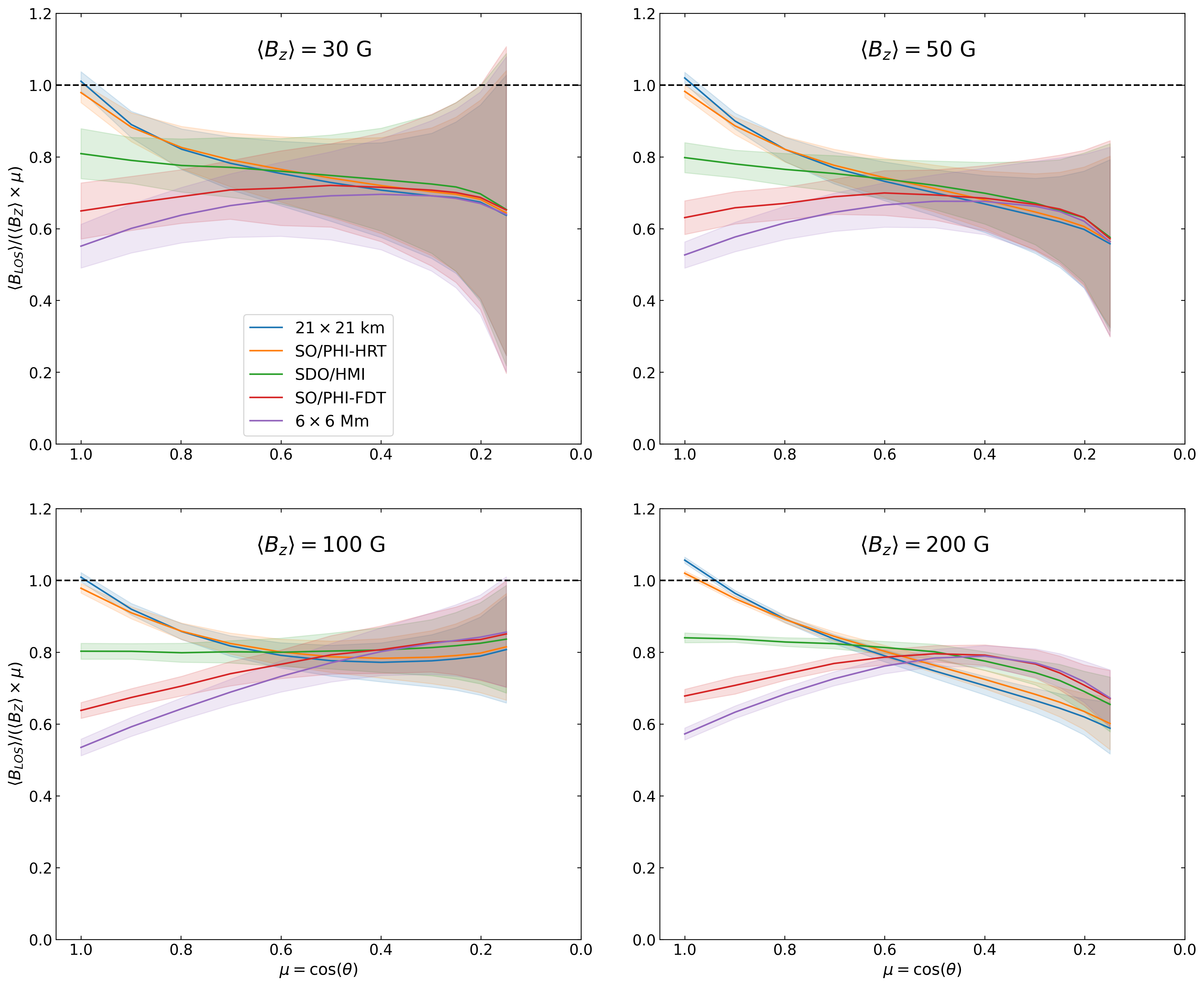}
  \caption{Same as Fig.~\ref{fig:mdi_blos_clv} but Fe~{\sc i}~$5250.2\;$\AA\;and $B_{LOS}$ is inferred by the MDI-like algorithm.} \label{fig:wfa_clv5250}%
\end{figure*}

\begin{figure*}[hb]
  \centering
  \includegraphics[width=\linewidth]{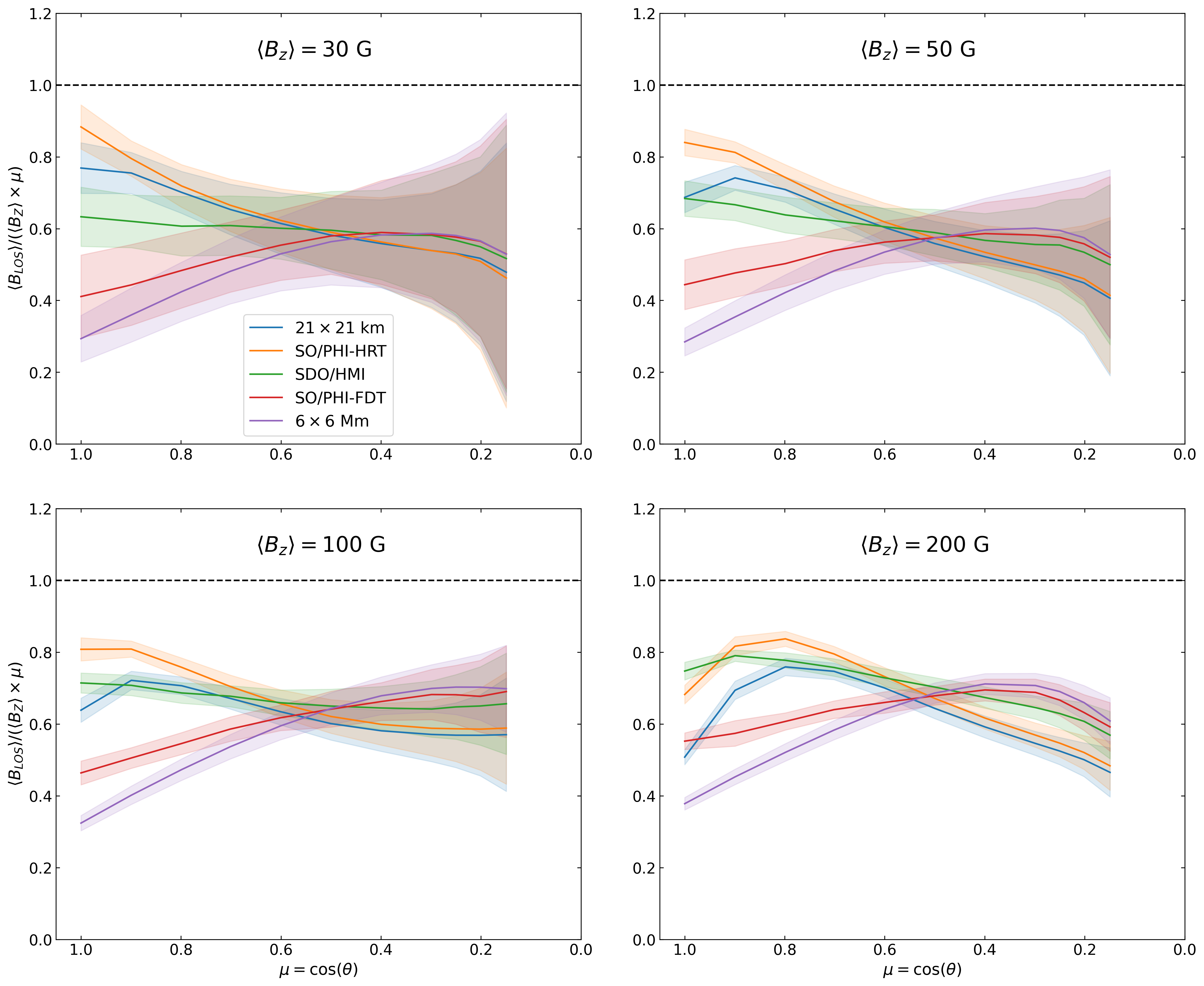}
  \caption{Same as Fig.~\ref{fig:blos_clv_frac} but Fe~{\sc i}~$5250.2\;$\AA\;and $B_{LOS}$ is inferred by the weak-field approximation.} \label{fig:wfa_clv5250}%
\end{figure*}

\begin{figure*}
  \centering
  \includegraphics[width=\linewidth]{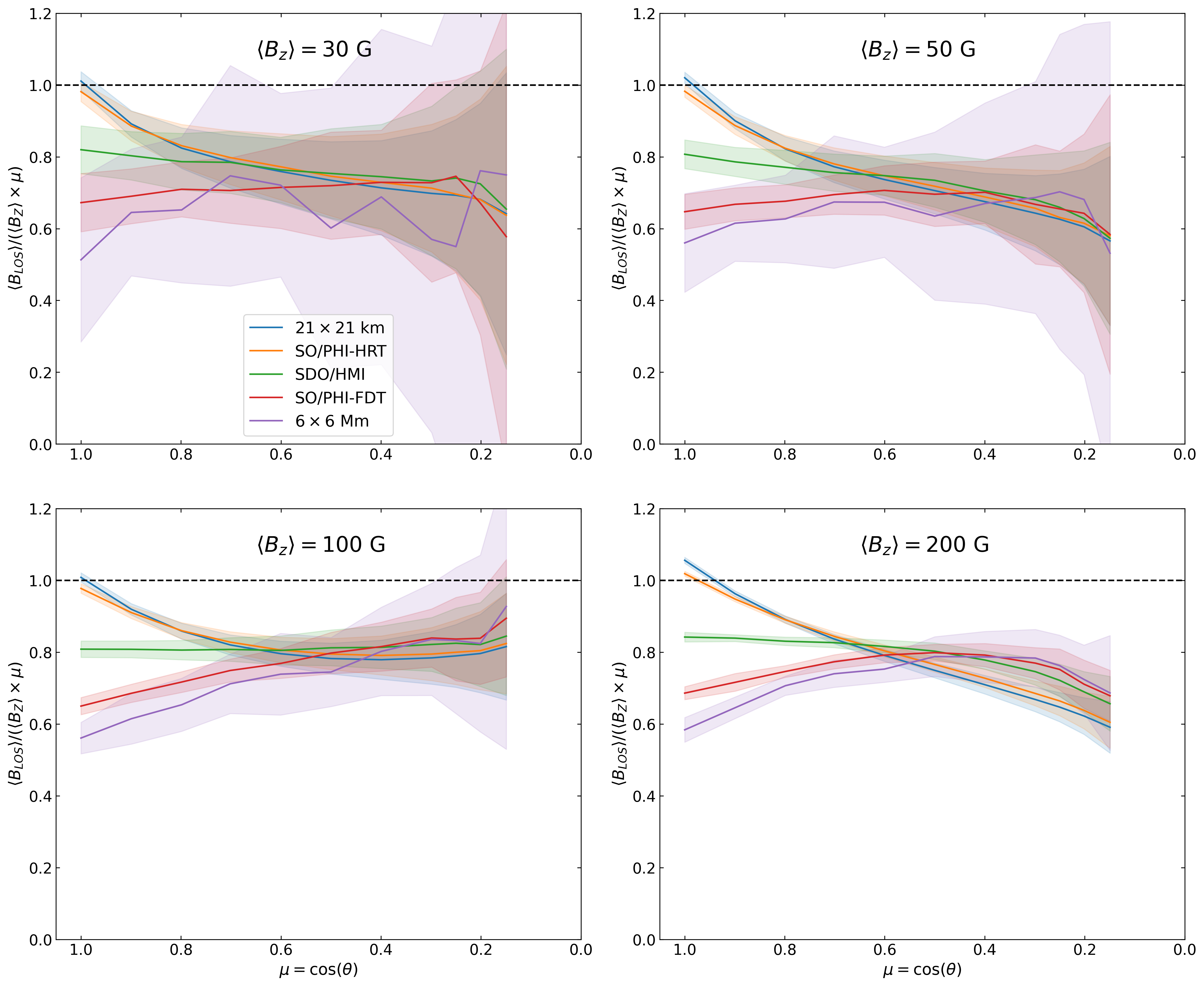}
  \caption{Same as Fig.~\ref{fig:blos_clv_frac} but Fe~{\sc i}~$5250.2\;$\AA\;and $B_{LOS}$ is inferred by the centre-of-gravity technique.} \label{fig:cog_clv5250}%
\end{figure*}

\end{document}